\documentclass[twoside]{ima} 
\usepackage{amssymb,amsmath,graphicx,epsfig}
\usepackage{mathptm} 
\usepackage{cmgreek} 
\usepackage{cite}
\usepackage[small,normal,bf,up]{caption2}
\usepackage{color}
\usepackage{multicol}

\newcommand{\blemma}{\begin{lemma}\rm}
\newcommand{\elemma}{\end{lemma}}
\newcommand{\bconjecture}{\begin{conjecture}\rm}
\newcommand{\econjecture}{\end{conjecture}}
\newcommand{\bdefinition}{\begin{definition}\rm}
\newcommand{\edefinition}{\end{definition}}
\newcommand{\bexample}{\begin{example}\rm}
\newcommand{\eexample}{\end{example}}

\newcommand{\expectation}{\ensuremath{\mathbb{E}}}

\newcommand{\QED}{\hspace{\fill}$\Box$}
\newcommand{\EEx}{\QED}
\newcommand{\dlmin}{{\tt l}_{\text{min}}} 
 
\newcommand{\dr}{{\tt r}} 
\newcommand{\dl}{{\tt l}}

\newcommand{\eldpc}[3]{\text{LDPC}(#1, #2, #3)}

\newcommand{\expeldpc}[4]{\text{ELDPC}(#1, #2, #3, #4)}

\newcommand{\graph}{{{\tt G}}} 

\newcommand{\parity}{{H}}

\newcommand{\hshannon}[2]{{\cal H}(#1, #2)}
\newcommand{\ledge}{\lambda}           
\newcommand{\redge}{\rho}              
\newcommand{\lnode}{\Lambda}           
\newcommand{\rnode}{P}                 

\newcommand{\xl}{x_{l}}

\newcommand{\xleft}{x_{l}}
\newcommand{\xright}{x_{r}}

\DeclareMathOperator{\prob}{{\mbox{\rm P}}}

\newcommand{\pth}[1]{\left( #1 \right)}

\newcommand{\Wh}{\widehat{W}}
\newcommand{\Xh}{\widehat{X}}
\newcommand{\Ph}{\widehat{P}}
\newcommand{\oX}{\overline{X}}
\newcommand{\oz}{\overline{z}}
\newcommand{\oth}{\overline{\theta}}
\newcommand{\fh}{\hat{f}}

\newcommand{\Wt}{\widetilde{W}}
\newcommand{\Hp}{{\mathbb H}_+}
\newcommand{\R}{{\mathbb R}}

\newcommand{\vD}{\vec{\Delta}}
\newcommand{\vu}{\vec{u}}
\newcommand{\vx}{\vec{x}}
\newcommand{\vX}{\vec{X}}
\newcommand{\vz}{\vec{z}}

\newcommand{\Ai}{{\rm Ai}}
\newcommand{\Bi}{{\rm Bi}}

\newcommand{\ve}{\varepsilon}

\begin{document}
\markboth{Finite-Length Scaling}{
	       Finite-Length Scaling}        

\title{Finite-Length Scaling for Iteratively Decoded LDPC Ensembles}

\author{
Abdelaziz Amraoui\thanks{EPFL (Lausanne), CH-1015, email:{\tt abdelaziz.amraoui@epfl.ch}}
\and
Andrea Montanari\thanks{LPTENS (UMR 8549, Unit{\'e}   Mixte de Recherche du
CNRS et de l' ENS), 24, rue Lhomond, 75231 Paris CEDEX 05, France, 
email:{\tt montanar@lpt.ens.fr}}
\and
Tom Richardson\thanks{Flarion Technologies, Bedminster, NJ,  USA-07921, email:{\tt richardson@flarion.com}}
\and
R{\"{u}}diger Urbanke\thanks{EPFL (Lausanne), CH-1015, email:{\tt rudiger.urbanke@epfl.ch}}
}

\maketitle   

\begin{abstract}
In this paper we investigate the behavior of iteratively decoded
low-density parity-check codes over the binary erasure channel in the
so-called ``waterfall region." We show that the performance curves in 
this region follow a very basic scaling law. We conjecture that 
essentially the same scaling behavior applies in a much more general
setting and we provide some empirical evidence to support this conjecture.
The scaling law, together with the error floor expressions developed previously,
can be used for fast finite-length optimization.
\end{abstract}


\newcommand{\mydeltat}{e}
\newcommand{\mydeltas}{l_0}
\newcommand{\mysigma}{l_1}
\newcommand{\mytau}{\dl-l_0-l_1}
\newcommand{\deltat}{\Delta t}
\newcommand{\deltas}{\Delta s}
\newcommand{\tdrift}{\text{f}^{(t)}}
\newcommand{\sdrift}{\text{f}^{(s)}}
\newcommand{\tvariance}{\text{f}^{(tt)}}
\newcommand{\svariance}{\text{f}^{(ss)}}
\newcommand{\tsvariance}{\text{f}^{(ts)}}

\section{Introduction}
\label{sec:introduction}

It is probably fair to say that the asymptotic behavior
(as the blocklength tends to infinity) of iterative 
coding systems is reasonably well understood to date. Much less
is known about the {\em finite}-length behavior though.

As usual, the situation is clearest for the binary erasure channel
(BEC$(\epsilon)$). In this case, the finite-length analysis of the
average performance of an ensemble boils down to
a combinatorial problem. In \cite{DPRTU01} recursions where
given to solve this combinatorial problem for 
some simple regular ensembles. These recursions
were generalized in \cite{ZaO02,RiU02} to deal with irregular ensembles,
expurgation and to compute block as well as bit erasure probabilities. Therefore,
in principle, by solving the corresponding recursions
it is possible to determine the average finite-length
performance for any desired ensemble. In practice
though this approach runs into computational
limitations. Roughly, the complexity of the
recursions grows by a factor $n$ (the blocklength) for
each degree of freedom of the ensemble. For reasonable
lengths therefore only very simple ensembles can
currently be analyzed in this way. 

Given the computational complexity of an exact finite-length
analysis, it is of great interest to find good approximations.
Let us consider ensembles whose threshold is not determined by
the stability condition, see \cite{LMSSS97}.
In this case, the finite-length performance curve can be
divided into two regions, \cite{RSU02}. The {\em waterfall} region
and the {\em error floor} region. In the waterfall region 
the performance is determined by `large' (linear sized) failures
and it improves quickly for
decreasing erasure probabilities. 
In the error floor region on the other hand
the performance is determined by `small' (sublinear sized)
weaknesses in the graph. Fortunately, this second region is relatively
easy to handle as was demonstrated in \cite{RSU02}.

In this paper we address the issue of modeling
the behavior of large error events. Our approach
is motivated by a general conjecture stemming from
statistical physics \cite{Fish71,Priv90}: If a system, parametrized by
lets say $\epsilon$, goes through a {\em phase
transition} at a critical parameter, call it $\epsilon^*$
(in our case the threshold), then it has repeatedly been
observed that around this critical parameter there
is a very specific scaling law. To be more concrete:
We are interested in the probability of block error 
as a function of the block length $n$ and the channel parameter 
$\epsilon$, call it $\prob_{\text{B}}(n, \epsilon)$.
We know that as $n$ tends to infinity there is 
a phase transition at $\epsilon^{*}$, the iterative decoding  
threshold. Asymptotically, $\prob_{\text{B}}(n, \epsilon)$ tends to zero
for $\epsilon < \epsilon^*$ and to one for $\epsilon > \epsilon^*$.
The scaling law refines this basic observation: One expects that 
there exists a non-negative constant $\nu$ 
and some non-negative function $f(z)$ so that
\begin{eqnarray}
\lim_{\overset{ n \rightarrow \infty}{{\rm s.t.} \ n^{1/\nu} \pth{\epsilon^*-\epsilon} = z}} 
\prob_{\text{B}}(n, \epsilon) = f(z).\label{equ:generalfss}
\end{eqnarray}

In other words, if one plots $\prob_{\text{B}}(n, \epsilon)$
as a function of $z=n^{\frac{1}{\nu}} (\epsilon^*-\epsilon)$
then, for increasing $n$ these finite-length curves
are expected to converge to some function $f(z)$. The function
$f(z)$ decreases smoothly from $1$ to $0$ as its argument changes from 
$-\infty$ to $+\infty$.
This means that  all finite-length curves are, to first order, 
scaled versions of some {\em mother} curve $f(z)$.
It might be helpful to think of the
threshold $\epsilon^{*}$ as the zero order term in
a Taylor series. Then the above scaling, if correct, represents
the first order term.
In fact, one can even refine the analysis to include higher order terms and write
\begin{eqnarray*}
\prob_{\text{B}}(n, \epsilon) & = & f(z) + n^{-\omega} g(z) + o(n^{-\omega}),
\end{eqnarray*}
where $\omega$ is some positive real number and $g(z)$ is the 
second order correction term.


Such scaling laws are expected to apply in a wide array of situations
in communications. The following is probably the simplest case in which
such a scaling law can be proven rigorously. Let $\hshannon n r$ denote Shannon's random
parity-check ensemble of codes of length $n$ and rate $r$. 
Consider transmission over the BEC$(\epsilon)$ using a 
random element of $\hshannon n r$ with maximum likelihood
(ML) decoding. Let $\parity$ denote a random parity-check matrix, 
let ${\cal E}$ denote the set of erased positions and let $\parity_{\cal E}$
denote the submatrix of $\parity$ consisting of the columns of $\parity$ indexed by 
${\cal E}$. The ML block decoder will succeed if and only if $\parity_{\cal E}$ has 
rank $E:=|{\cal E}|$. By definition, $\parity_{\cal E}$ is itself a random binary matrix of
dimension $E \times n \bar{r}$, where $\bar{r}:=1-r$. Some thought shows that 
\[
\prob \left\{ \text{rank} \left( \parity_{\cal E} \right) = E \right\}  = 
\begin{cases}
0, & E > n \bar{r}, \\
\prod_{i=0}^{E-1} \left(1-2^{i-n \bar{r}} \right), & 0 \leq E \leq n \bar{r}.
\end{cases}
\]
A quick calculation reveals that
\begin{eqnarray*}
& & \expectation_{\hshannon n r}[\prob_{\text{B}}(\parity, \epsilon)] \\
& = & 
\sum_{E=0}^{n\bar{r}} \binom{n}{E} \epsilon^E \bar{\epsilon}^{n-E} 
\left(1-\prod_{i=0}^{E-1} \left(1-2^{i-n\bar{r}} \right) \right) + \sum_{E=n\bar{r}+1}^n \binom{n}{E} \epsilon^E \bar{\epsilon}^{n-E} \\
& = & Q \left(\frac{ \sqrt{n}(\epsilon^*-\epsilon)}{\sqrt{\epsilon^* \bar{\epsilon^*}}} \right) \left( 1+O(1/n) \right),
\label{equ:hshannonscaling}
\end{eqnarray*}
where in the last line we used the fact that $\bar{r}=\epsilon^*$
and we defined the $Q$-function as usual by 
\[
Q(z) :=\frac{1}{\sqrt{2\pi}}\int_{z}^{\infty}\!\!\!e^{-x^2/2}\;dx\, .
\]
In words, since the conditional probability of block erasure falls off steeply away from the
threshold, the scaling law is dominated by the probability that the channel behaves
atypically and that the number of erasures exceeds $n \epsilon^*=n \bar{r}$.

In this paper we prove a scaling law for iteratively decoded standard ensembles $\eldpc n {\ledge} {\redge}$
and Poisson ensembles $\eldpc n {\ledge} {r}$ 
when transmission takes place over the BEC$(\epsilon)$.
In the sequel we give a leisurely overview regarding the main results. 
The precise statements can be found in Section \ref{sec:general}. 
Some of the background material is summarized in Section \ref{sec:review}.

Assume first that $\dlmin \geq 3$, i.e., that the minimum left degree is at least three.
Let $\graph$ be a random element of the ensemble.
Then, as stated more precisely in Section \ref{sec:general},  
\begin{eqnarray}
\expectation[\prob_{\text{B}}(\graph, \epsilon)] = Q \left( \frac{ \sqrt{n} (\epsilon^*-\epsilon)}{\alpha} \right)+o(1),\label{equ:simplscaling}
\end{eqnarray}
where $\alpha$ is a quantity which depends on the ensemble and
which is computable by a procedure similar to density evolution.
This scaling law has a form almost identical to (\ref{equ:hshannonscaling}) with
$\alpha^2$ representing a variance. Therefore we dub the procedure which leads to
the computation of $\alpha$, {\em covariance evolution}.
We conjecture that in fact the following refined scaling law is valid,
\begin{eqnarray} 
\expectation[\prob_{\text{B}}(\graph, \epsilon)] & = & Q\left( \frac{ \sqrt{n} (\epsilon^*-\epsilon)}{ \alpha} \right) + 
\beta n^{-\frac{1}{6}}\frac{1}{\sqrt{2 \pi \alpha^2}} e^{-\frac{n (\epsilon^*-\epsilon)^2 }{2 \alpha^2}} +O(n^{-1/3})\nonumber \\
& = & Q\left(\frac{\sqrt{n} (\epsilon^*-\beta n^{-\frac{2}{3}}-\epsilon)}{ \alpha} \right)+O(n^{-1/3}), \label{equ:scaling}
\end{eqnarray}
where the term $\beta n^{-\frac{2}{3}}$ represents a {\em shift} of
the threshold for finite lengths. Again, this constant $\beta$ depends on the ensemble and
we will show how it can be computed. 

\begin{figure}[htp]
\begin{center}
\setlength{\unitlength}{0.8bp}%
\begin{picture}(0,0)
\includegraphics[scale=0.8]{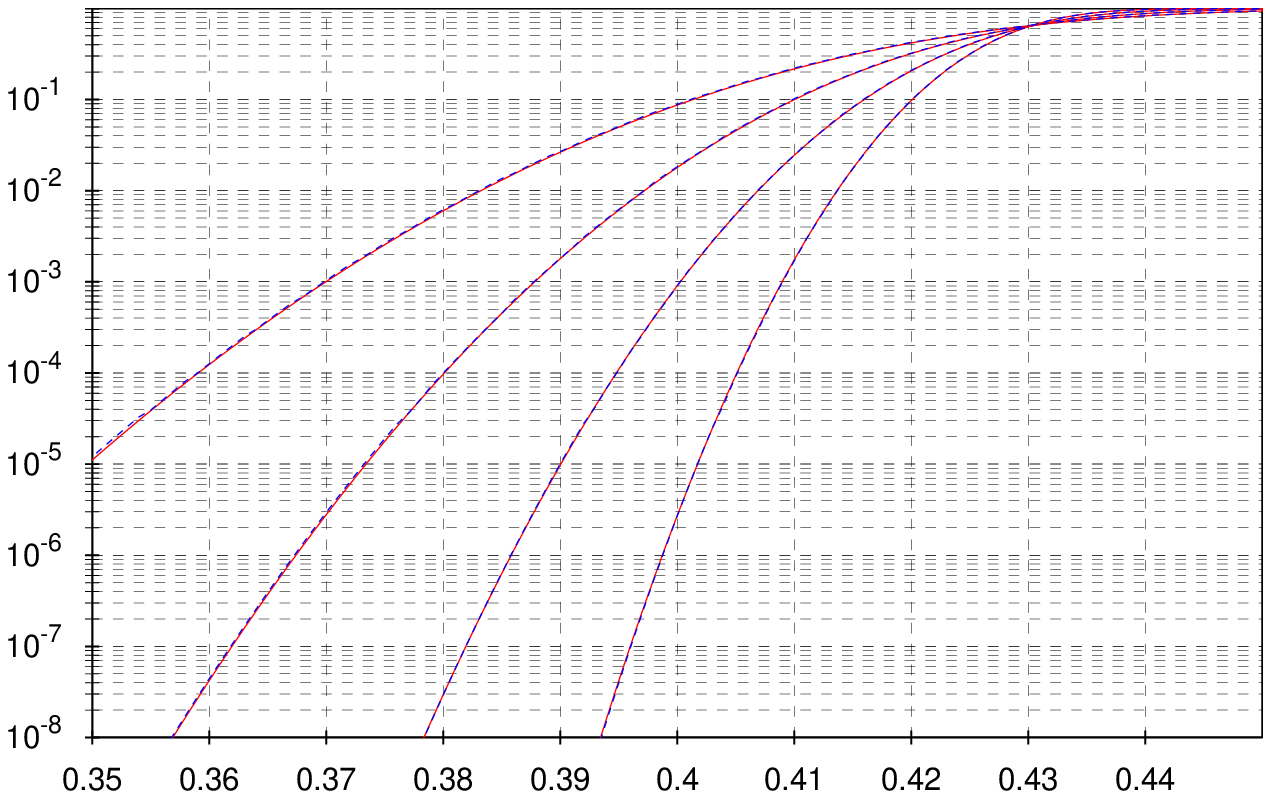}
\end{picture}%
\begin{picture}(367.5,231)
\put(13,223){\makebox(0,0){$\prob_{\text{B}}$}}
\put(366,8){\makebox(0,0)[r]{$\epsilon$}}
\end{picture}
\end{center}
\caption{\label{fig:scalingbec}
Scaling of $\expectation_{\eldpc n {x^2} {x^5}}[\prob_{\text{B}}(\graph, \epsilon)]$ for
transmission over BEC$(\epsilon)$ and belief propagation decoding.
The threshold for this combination is $\epsilon^* \approx
 0.42944$, see Table \ref{tab:ParametersRegular}.
The blocklengths/expurgation parameters are 
$n/s=1024/24$, $2048/43$, $4096/82$ and $8192/147$, respectively.
(More precisely, we assume that the ensembles
have been expurgated so that graphs in this ensemble do not contain stopping sets of size $s$ or smaller.)
The solid curves represent the exact ensemble averages.
The dashed curves are computed according to the
refined scaling law stated in Conjecture \ref{RefinedConjecture} with scaling parameters 
$\alpha=\sqrt{0.249869^2+\epsilon^*(1-\epsilon^*)}$ 
and $\beta=0.616045$, see Table \ref{tab:ParametersRegular}.}
\end{figure}

Figure \ref{fig:scalingbec} shows this scaling applied to the $\eldpc n {x^2} {x^5}$ ensemble
which will serve as our running example.
Note that the above scaling law models the
behavior of {\em large} error events. A better comparison with equation 
(\ref{equ:scaling}) is therefore obtained by considering
{\em expurgated} ensembles, see \cite{RSU02}. For $\dlmin \geq 3$ the scaling (\ref{equ:scaling}) 
holds true asymptotically regardless of the expurgation scheme. 
This follows since, as shown in \cite{ZaO02}, the contribution to the block
error probability stemming from sublinear-sized weaknesses 
in the graph decreases like\footnote{In the sequel we follow the standard
convention to write $O(\cdot)$ to denote an {\em upper bound}
but we write $\Theta(\cdot)$ to denote the exact behavior (up to constants).}
$\Theta \left(n^{1-\lceil \dlmin/2 \rceil}\right)$.
This is the probability of having a stopping set formed by a single
variable node and $\lfloor \dlmin/2 \rfloor$ check nodes (such a constellation
is allowed unless double edges are forbidden).

The situation is somewhat more complicated once $\lambda'(0)>0$. In this case
the block erasure probability consists of two parts: the part which stems from
linear-sized error events and which scales like (\ref{equ:scaling})
and a contribution which stems from sub-linear sized weaknesses in the graph.
The contribution from the latter part depends crucially on the expurgation 
scheme employed and does not necessarily vanish as $n\to\infty$.

\begin{figure}[htp]
\begin{center}
\setlength{\unitlength}{0.8bp}%
\begin{picture}(0,0)
\includegraphics[scale=0.8]{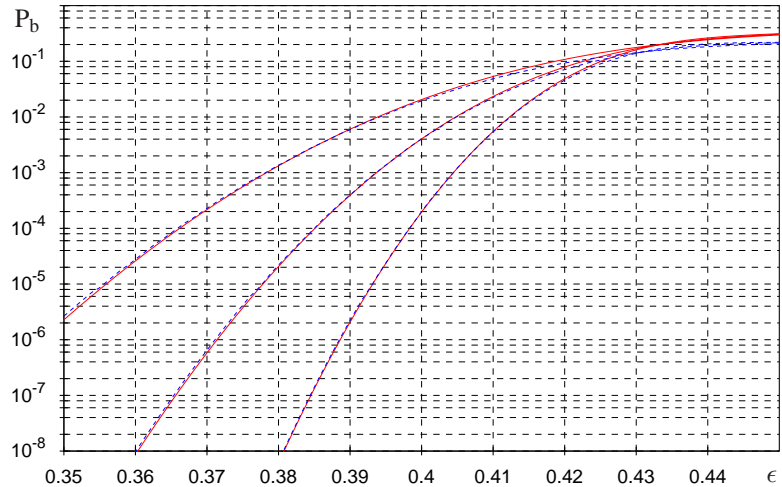}
\end{picture}%
\begin{picture}(367.5,231)
\put(13,223){\makebox(0,0){$\prob_{\text{b}}$}}
\put(366,8){\makebox(0,0)[r]{$\epsilon$}}
\end{picture}
\end{center}
\caption{\label{fig:scalingbecbit}
Scaling of $\expectation_{\eldpc n {x^2} {x^5}}[\prob_{\text{b}}(\graph, \epsilon)]$ for
transmission over BEC$(\epsilon)$ and belief propagation decoding.
The threshold for this combination is $\epsilon^* \approx 0.42944$, see Table \ref{tab:ParametersRegular}.
The blocklengths/expurgation parameters are
$n/s=1024/24$, $2048/43$ and $4096/82$,
respectively.
The solid curves represent the exact ensemble averages.
The dashed curves are computed according to the
refined scaling law stated in Conjecture \ref{RefinedConjecture} with scaling parameters
$\alpha=\sqrt{0.249869^2+\epsilon^*(1-\epsilon^*)}$
and $\beta=0.616045$, see Table \ref{tab:ParametersRegular}.
}
\end{figure}

In the above discussion we focused on the {\em block} erasure probability.
The equivalent scaling law for the bit erasure probability is a straightforward
adaptation: If the decoder fails at the critical\footnote{See Section \ref{sec:review}
for a discussion of this notion.} point then, asymptotically, it incurs
a fixed bit erasure probability, call it $\nu^*$ (the fractional size of
the residual graph). Therefore, if we multiply the above
expressions by $\nu^*$ we get the corresponding scaling law for the bit erasure probability.\footnote{
The approximation can be improved away from the threshold by 
multiplying the above expression with the typical size of the failure for that particular $\epsilon$.}
Figure \ref{fig:scalingbecbit} shows the resulting approximation of
$\expectation_{\eldpc n {x^2} {x^5}}[\prob_{\text{b}}(\graph, \epsilon)]$. 

\begin{figure}[htp]
\begin{center}
\setlength{\unitlength}{0.8bp}%
\begin{picture}(0,0)
\includegraphics[scale=0.8]{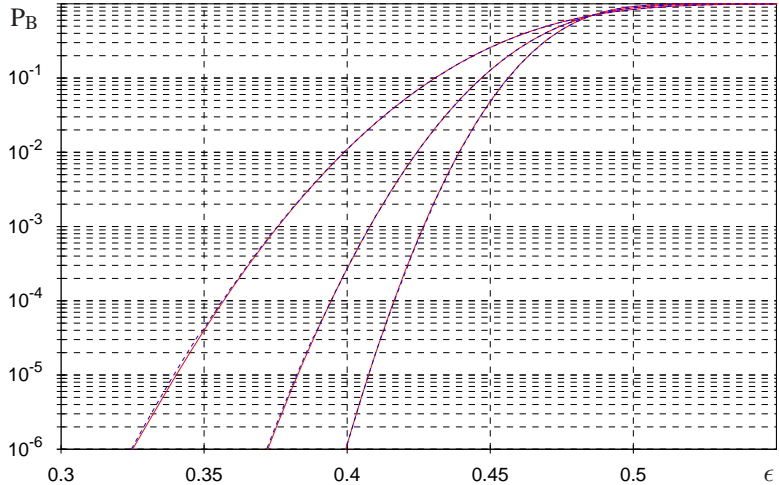}
\end{picture}%
\begin{picture}(367.5,231)
\put(13,223){\makebox(0,0){$\prob_{\text{B}}$}}
\put(366,8){\makebox(0,0)[r]{$\epsilon$}}
\end{picture}
\end{center}
\caption{\label{fig:irregular}
Scaling of $\expectation_{\eldpc n {\ledge=\frac{1}{6}x+\frac{5}{6}x^3} {\redge=x^5}}[\prob_{\text{B}}(\graph, \epsilon)]$ for
transmission over BEC$(\epsilon)$ and belief propagation decoding.
The threshold for this combination is $\epsilon^* \approx 0.48281$.
The blocklengths/expurgation parameters are $n/s=350/14$, $700/23$ and $1225/35$. 
The solid curves represent the simulated ensemble averages.
The dashed curves are computed according to the
refined scaling law stated in Conjecture \ref{RefinedConjecture} with scaling parameters
$\alpha=\sqrt{0.276^2+\epsilon^*(1-\epsilon^*)}$
and $\beta=0.642274$. These parameters were fitted to the data.}
\end{figure}

The basic form of the scaling law applies to regular as well as irregular ensembles.\footnote{This
is true as long as the threshold is not determined by the stability condition and is determined by
a single critical point, see Sections \ref{sec:review} and \ref{sec:general}.}
The computation of the scaling parameters though becomes significantly more
involved in the irregular case and therefore we limit ourselves in this paper 
to providing the detailed calculations only for regular ensembles.
Fig.~\ref{fig:irregular} demonstrates the scaling law for the block erasure probability
applied to the irregular ensemble $\eldpc n {\ledge=\frac{1}{6}x+\frac{5}{6}x^3}
{\redge=x^5}$. In this case the scaling parameters were simply fitted 
to the data.

\begin{figure}[htp]
\begin{center}
\setlength{\unitlength}{0.8bp}%
\begin{picture}(0,0)
\includegraphics[scale=0.8]{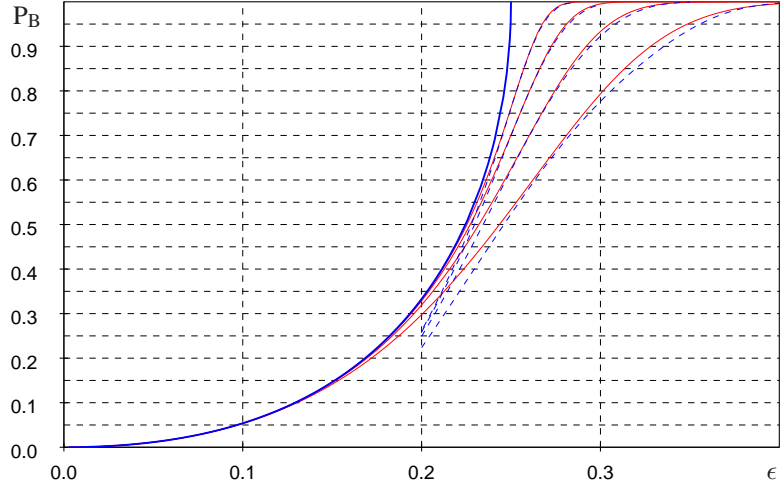}
\end{picture}%
\begin{picture}(367.5,231)
\put(13,223){\makebox(0,0){$\prob_{\text{B}}$}}
\put(366,8){\makebox(0,0)[r]{$\epsilon$}}
\end{picture}
\end{center}
\caption{\label{fig:poisson}
Scaling of $\expectation_{\eldpc n {\ledge=x} {r=\frac{1}{2}}}[\prob_{\text{B}}(\graph, \epsilon)]$
for transmission over the BEC$(\epsilon)$ and belief propagation decoding. The (bit) threshold for this
combination is $\epsilon^* = \frac{1}{4}$. The solid curves are the exact ensemble averages for blocklengths equal to
$n=256$, $1024$, $4096$ and $16384$. The bold curve is the limiting (in $n$) block erasure curve. The dashed curves are the
finite-length approximations computed according to equation (\ref{BlockScaling}).}
\end{figure}

The performance of ensembles whose threshold is determined by the stability condition
scales in a fundamentally different way. 
The simplest such representatives are cycle codes. We will discuss cycle codes
in some detail since we conjecture that the same scaling applies to all ensembles
for which the stability condition determines the threshold.
Fig.~\ref{fig:poisson} shows block erasure curves for the
$\eldpc n {x} {r=\frac{1}{2}}$ cycle Poisson ensemble with expurgation parameter $s=1$ for
$n=2^i$, $i=8, 10, 12, 14$. 
Also shown is the limiting block erasure probability curves 
{\em and our approximation for the block error probability around the threshold}. Clearly, these curves
differ in their nature significantly from the curves discussed before.
As investigated in more detail in Section \ref{sec:general},
the block erasure probability
does not show a threshold effect: instead it converges to a smooth limiting 
curve.  Around the threshold we have the following scaling law, 
\begin{eqnarray}
\expectation_{\eldpc n {x} {r}}[\prob_{\text{B}}(\graph, \epsilon)] = 1- A a n^{-1/6}
\,  f(b\, n^{1/3}(\epsilon-\epsilon^*))\,\left\{
1+O(n^{-1/3})\right\}\, ,\nonumber\\
\label{BlockScaling}
\end{eqnarray}
where $a = \bar{r}^{-1/6}$, $b = \bar{r}^{-2/3}$ and $A$ is a constant which
depends on the expurgation scheme used. The form of the mother curve $f(x)$
is given in Lemma \ref{lem:blockpoisson}.

\subsection{Scaling for General Channels}
\begin{figure}[htp]
\begin{center}
\setlength{\unitlength}{0.8bp}%
\begin{picture}(0,0)
\includegraphics[scale=0.8]{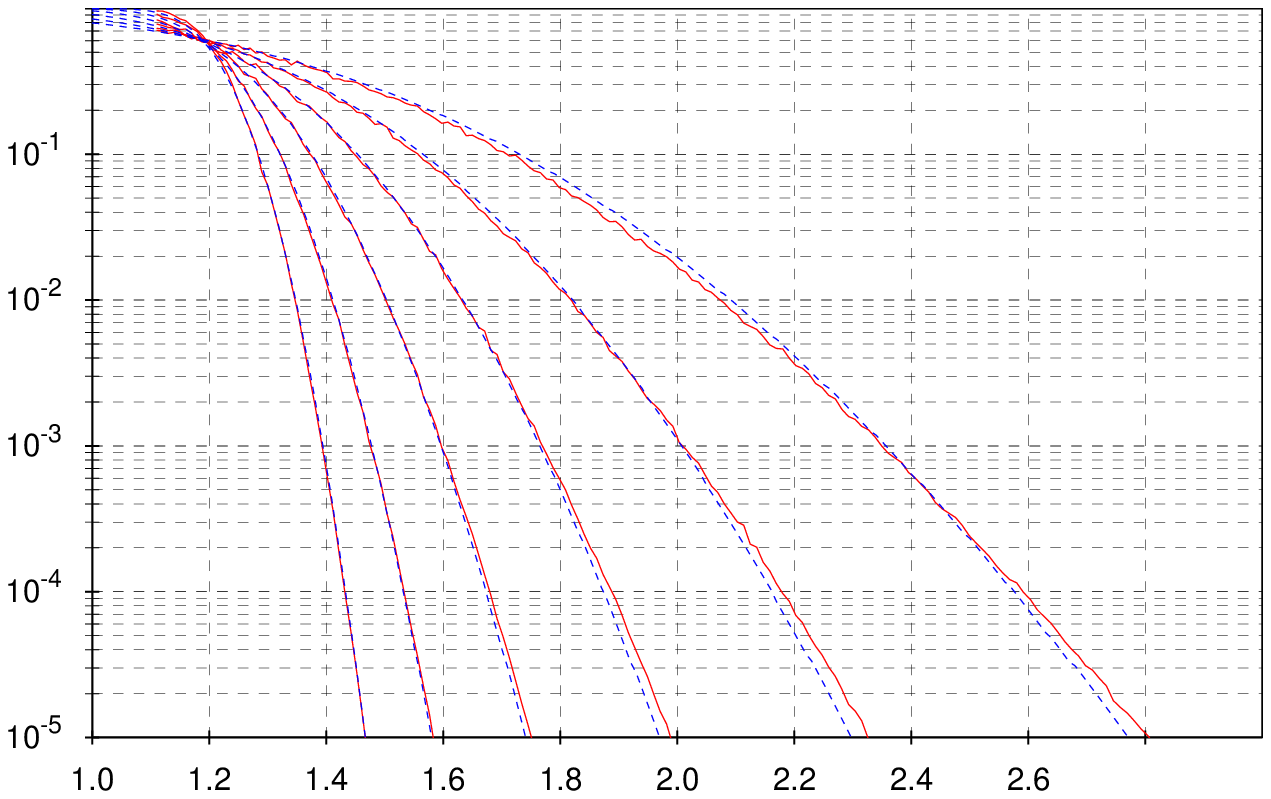}
\end{picture}%
\begin{picture}(367.5,231)
\put(13,223){\makebox(0,0){$\prob_{\text{B}}$}}
\put(366,8){\makebox(0,0)[r]{$\left( E_b/N_0 \right)_{\text{dB}}$}}
\end{picture}
\end{center}
\caption{\label{fig:scalingawgnquant}
Scaling of $\expectation_{\eldpc n {x^2} {x^5}}[\prob_{\text{B}}(\graph, \sigma)]$ for
transmission over BAWGNC$(\sigma)$ and a quantized version of belief propagation decoding implemented in
hardware.  The threshold for this combination is 
$\left(E_b/N_0 \right)^*_{\text{dB}} \approx 1.19658$.
The blocklengths $n$ are
$n=1000$, $2000$, $4000$, $8000$, $16000$ and $32000$, respectively.
The solid curves represent the simulated ensemble averages.
The dashed curves are computed according to the
refined scaling law (\ref{equ:scaling}) with scaling parameters
$\alpha=0.8694$ and $\beta=5.884$. These parameters were fitted to the empirical data.}
\end{figure}

\begin{figure}[htp]
\begin{center}
\setlength{\unitlength}{0.8bp}%
\begin{picture}(0,0)
\includegraphics[scale=0.8]{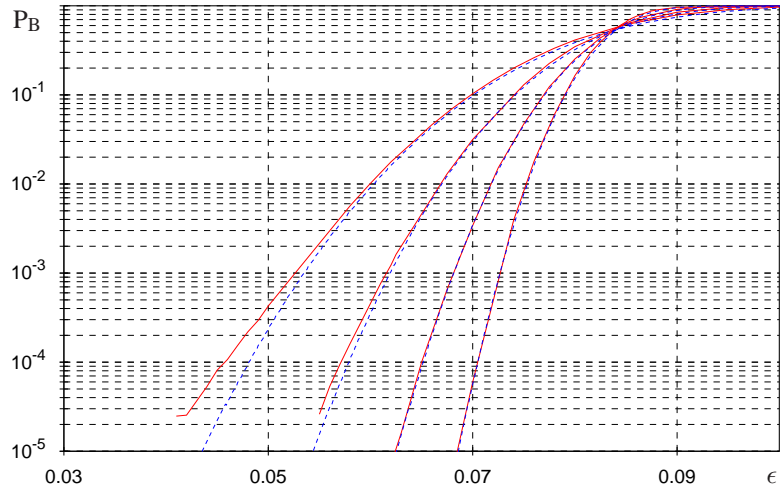}
\end{picture}%
\begin{picture}(367.5,231)
\put(13,223){\makebox(0,0){$\prob_{\text{B}}$}}
\put(366,8){\makebox(0,0)[r]{$\epsilon$}}
\end{picture}
\end{center}
\caption{\label{fig:scalingbsc}
Scaling of $\expectation_{\eldpc n {x^2} {x^5}}[\prob_{\text{B}}(\graph, \epsilon)]$ for
transmission over BSC$(\epsilon)$ and belief propagation decoding.
The threshold for this combination is $\epsilon^* \approx 0.084$.
The blocklengths/expurgation parameters are $n/s=1024/19$, $2048/39$, $4096/79$ and $8192/79$, respectively. 
The solid curves represent the 
ensemble averages obtained via simulation.
The dashed curves are computed according to the
refined scaling law stated in equation (\ref{equ:scaling}) with scaling parameters $\alpha=1.156$ and $\beta=0.1$.} 
\end{figure}

\begin{figure}[htp]
\begin{center}
\setlength{\unitlength}{0.8bp}%
\begin{picture}(0,0)
\includegraphics[scale=0.8]{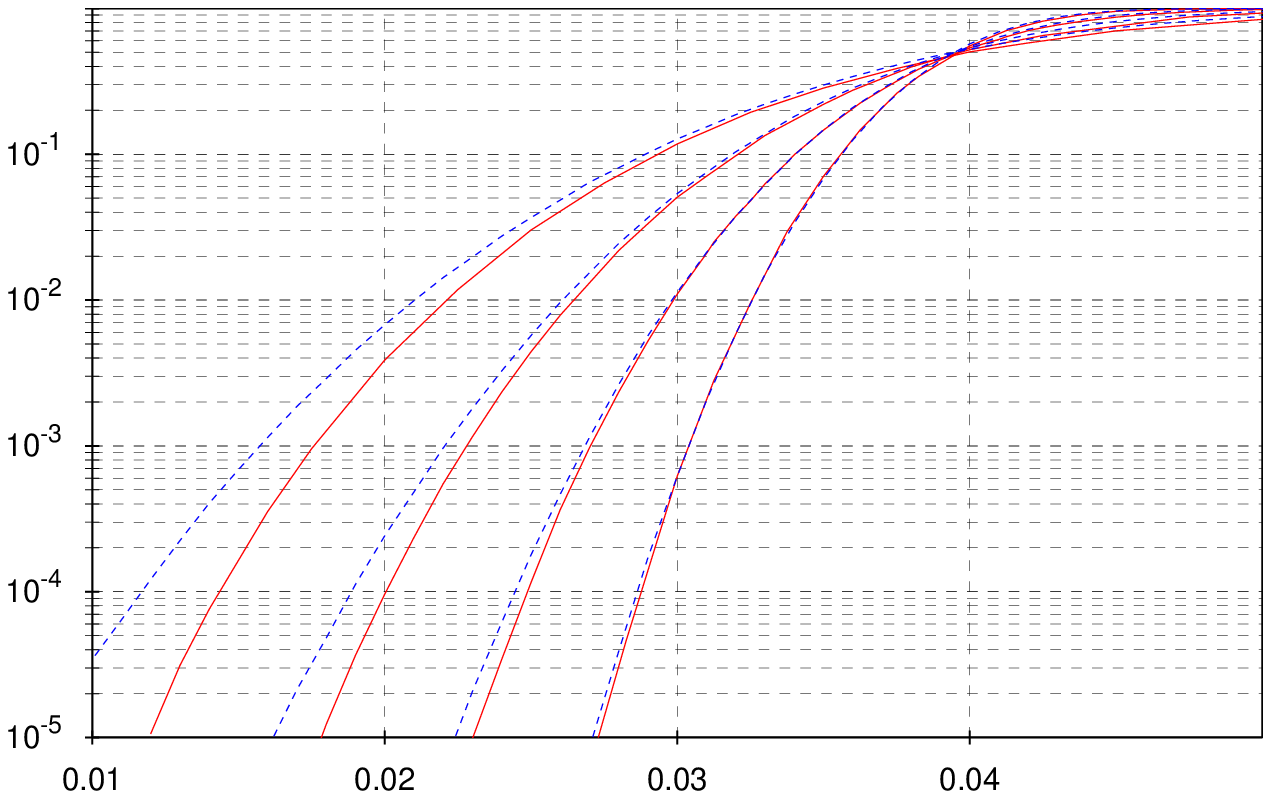}
\end{picture}%
\begin{picture}(367.5,231)
\put(13,223){\makebox(0,0){$\prob_{\text{B}}$}}
\put(366,8){\makebox(0,0)[r]{$\epsilon$}}
\end{picture}
\end{center}
\caption{\label{fig:scalinggal}
Scaling of $\expectation_{\eldpc n {x^2} {x^5}}[\prob_{\text{B}}(\graph, \epsilon)]$ for
transmission over BSC$(\epsilon)$ and Gallager Algorithm A decoding.
The threshold for this combination is $\epsilon^* \approx 0.03946$.
The blocklengths/expurgation parameters are $n/s=512/50$, $1024/70$, $2048/100$ and $4096/200$, respectively.
The solid curves represent the ensemble averages obtained via simulation.
The dashed curves are computed according to the
refined scaling law stated in equation (\ref{equ:scaling}) with scaling parameters $\alpha=1.11$ and $\beta=0.0$.}
\end{figure}

In many ways this paper only represents the very first step in what seems to
be a promising research direction.
The most important extension is undoubtedly the one to general binary-input
output-symmetric channels. 
Although there is currently little hope of
attacking this problem rigorously, empirically such a scaling seems
to be true for general channels as well. 
In principle any (function of the) channel parameter 
can be used for stating the scaling law, however we make this
choice slightly less arbitrary by the following convention.
Consider a family of binary-input output-symmetric memoryless channels
parametrized by lets say $\sigma$. Let $C(\sigma)$ denote the capacity
for the parameter $\sigma$. The role of $\epsilon^*-\epsilon$ in the case of the BEC$(\epsilon)$
is then played by $C(\sigma)-C(\sigma^*)$, i.e., we use the scaling law
\begin{equation}
\label{equ:generalscaling}
P_B = Q \left( \frac{\sqrt{n}(C(\sigma)-C(\sigma^*)-\beta n^{-\frac{2}{3}})}{\alpha} \right).
\end{equation}
Note that for the BEC$(\epsilon)$, $C(\epsilon)=1-\epsilon$, so that 
this choice is consistent with our previous convention.
The parameters $\alpha$ and $\beta$ reported in the captions of
Figs.~\ref{fig:scalingawgnquant} to \ref{fig:scalinggal} are defined
according to the above formula.

Fig.~\ref{fig:scalingawgnquant} shows performance curves for the $\eldpc n {\ledge=x^2} {\redge=x^5}$ ensemble
transmitted over the binary-input additive white Gaussian noise (BAWGN) channel and a quantized
version of belief propagation. Fig.~\ref{fig:scalingbsc} shows the corresponding
curves for the same ensemble when transmission takes place over the 
binary symmetric channel (BSC) and belief propagation decoding is used.
Finally, Fig.~\ref{fig:scalinggal} shows the performance curve for the Gallager algorithm A.
Although these cases are quite distinct one can see that the empirically fitted
scaling laws are in excellent agreement with the exact curves.

\subsection{Applications of Scaling to Finite-Length Optimization}
\label{sec:extensions}
An important application of the scaling laws which is left for future work
is finite-length optimization. Combined with analytic expressions of
the contribution to the error probability stemming from small (sublinear sized) 
weaknesses of the graph, the scaling laws can be used as an approximation 
to the performance for finite-length ensembles. 
Note also that from the limited examples exhibited in this paper it appears 
that the scaling parameters depend only weakly on the degree distribution. 
This suggest that a good optimization strategy for finite-length ensembles
is to optimize the infinity threshold under the condition that 
the contribution of the error floor leads to acceptable overall performance. 

\subsection{Connected Work and Outline}
In \cite{LeBICC03} an approach to analyze the finite-length behavior of turbo-codes 
was introduced. This method, which the author call the ``Exit band chart'', is used to 
describe the probabilistic convergence of the iterative decoding algorithm and provides 
an approximation of the BER in the waterfall region. Somewhat related is also the work by Zemor
and Cohen 
who study in \cite{ZeC95} the ``threshold'' behavior of general classes of codes.
A preliminary numerical investigation of the scaling 
(\ref{equ:simplscaling}) was presented in \cite{Mon01}. Partial accounts
of the present work appeared in \cite{AMRU03, AMRU04}.

In Section \ref{sec:review} we introduce the necessary notation and review some
of the background material, in particular the density evolution analysis as
introduced by Luby et. al. in \cite{LMSSS97}. In Section \ref{sec:general} we 
state and prove the general form of the scaling laws. 
In Section \ref{sec:parameters} we then discuss for regular ensembles how
the scaling parameters can be computed. In section \ref{sec:shiftparameters} we discuss in detail the refined scaling law and how the shift parameter can be computed. Some of the background material
and some detailed calculations have been relegated to Appendices.

\section{Review}
\label{sec:review}
In this section we recall some basic facts on the density evolution
analysis of low-density parity-check (LDPC) codes under iterative decoding.
We also fix some of the notation to be used throughout the paper.

\subsection{Ensembles and Channel Models}

In this paper we consider both standard as well as 
Poisson low-density parity-check ensembles.
Standard ensembles are denoted in the usual way as 
$\eldpc n \ledge \redge$, where $n$ is the block length and $\ledge$ and $\redge$ denote
the degree distributions from an edge perspective, see \cite{LMSSS97}.
For the Poisson ensemble the
right degree distribution is Poisson. More precisely, given the left
degree distribution $\ledge$ and the {\em rate} $r$, the right degree 
distribution tends to $\rho(x)=e^{\frac{x-1}{\bar{r} \int \! \ledge}}$
as $n\to\infty$.
We will denote such an ensemble by $\eldpc n \ledge r$.
To sample from the Poisson ensemble pick a bipartite graph with $n$ variable nodes
and the proper variable node degree distribution. Connect each edge emanating 
from a variable node to one of the $n \bar{r}$ check nodes, where the choice is taken according
to a uniform probability distribution.

From time to time it is more convenient to describe the degree distributions
from a node perspective. Our notation for the left and right node degree
distributions are $\lnode$ and $\rnode$ respectively and we have the following
important relationships.
\[
\ledge(1)=\redge(1)=1; \;\;\; \lnode(1)=n, \rnode(1)=n \bar{r}.
\]

It will sometimes be necessary to consider expurgated ensembles. Although there
are many expurgation mechanisms possible, we will limit our discussion to the following
simple scheme. Consider e.g. the case of expurgated Poisson ensembles.
Define $\expeldpc n {\ledge} {r} {s}$ as the subset of all elements in 
$\eldpc n \ledge r$ whose minimum stopping set size is at least $s+1$.
As always, endow this
set with the uniform probability distribution.
E.g., $\expeldpc n {\ledge} {r} {2}$ denotes the Poisson
ensemble which contains no stopping sets of size one or two.
The same notational convention is used for expurgated standard ensembles.

We will consider two channel models. The more familiar one is the binary erasure channel
with parameter $\epsilon$, denoted by BEC$(\epsilon)$, where each bit is erased independently
with probability $\epsilon$. Sometimes though it is more convenient to consider the
model BEC$(n, n \epsilon)$, the channel model in which {\em exactly} $n \epsilon$ out
of all $n$ bits are erased and where the set of these $n \epsilon$ erased bits
is chosen uniformly from all $\binom{n}{n \epsilon}$ such choices. 

We consider scaling laws for both bit as well as block erasure probabilities and we will
always consider ensemble averages. E.g., in its full notational glory,
\[ 
\expectation_{\expeldpc n {\ledge(x)=x} {r=\frac{1}{2}} {s=1}}[\prob_{\text{B}}(\graph, n \epsilon)]
\]
will denote the expected block erasure probability for cycle Poisson ensembles of
rate one-half containing no double edges when transmitted over the channel BEC$(n, n \epsilon)$.
Because of the obvious notational burden we will often replace this with shorthands
and we might write e.g., 
\[
\prob_{\text{B}}(n, \ledge(x)=x, r = \frac{1}{2}, s=1, n \epsilon).
\]
We might even omit some of the parameters if they are 
clear from the context.

\subsection{Decoding}
There are essentially two alternative ways of defining the  
decoding algorithm for the BEC$(\epsilon)$. 
Although they are equivalent in performance they are quite different from
the point of view of analysis. 
First, we can think of the standard message passing decoder in which messages are passed
in parallel from left to right and then back from right to left until the
codeword has been decoded or no further progress is achieved, \cite{KFL01}.
Alternatively one can think of the decoder as a process which tries to 
determine one bit at a time in a greedy fashion. This is the point of view
introduced by Luby et al. in \cite{LMSSS97,LMSS01} and we will
adopt it  in this paper. More precisely, the decoder proceeds as follows.
Given the received message, the decoder passes all {\em known} values
on to the check node side. These values are accumulated at the check nodes
and this partial metric is stored.
Further, all known nodes and edges over which
messages have been passed are deleted. In this way one arrives at a {\em residual} graph
which has a certain degree distribution. The decoder proceeds now in an iterative
fashion. If the residual graph contains no degree-one check nodes the decoding process stops.
Otherwise, the decoder randomly choses one such degree-one check node
and passes its partial metric to the connected 
variable node. This variable node is now known. Its value is communicated to
all connected check nodes, where the value is
accumulated to the partial metric. The involved variable node, check node
and all involved edges are deleted. In this way a new residual graph results
and a new iteration starts.

\subsection{Density Evolution}
The advantage of the
second description lies in the fact that the decoding process is
seen as a stochastic process with small increments -- at each iteration the
change of the degree distribution is a random variable and this change
is small. By standard arguments one can show
that in the large blocklength limit the behavior of individual instances
follows with high probability the expected such behavior and
this expected behavior can be expressed as the solution of
a differential equation. This is the idea introduced in \cite{LMSSS97}.

First recall that by definition of the ensemble the degree
distribution of the residual graph constitutes a
{\em sufficient statistics}, i.e., given
this degree distribution all residual graphs which are compatible with this
degree distribution (and are compatible with the general description of
the ensemble, like, e.g., the degree of expurgation) are equally likely. 
Therefore, in order to analyze the behavior of the decoder it
suffices to analyze the evolution of this degree distribution.
Let us now recall the solution of the infinite length analysis given
in \cite{LMSSS97} since
it forms the starting point for our investigation.
Let $\xl$ denote the fraction of erasure messages entering the variable nodes
at a given point in time (here the $l$ stands for right-to-{\em left} message). In terms of this parametrization, the evolution 
of the system (i.e., the evolution of the degree distribution of
the residual graph) is given by 
\begin{eqnarray}
L_i\left(\xl \right) & = & \epsilon \lnode_i \xl^i, \;\; i \geq 2, \nonumber \\
R_0\left(\xl \right) & = & \rnode(1)-\sum_{j \geq 1} R_j\left(\xl \right), \nonumber \\
R_1\left(\xl \right) & = & \lnode'(1) \epsilon \ledge\left(\xl \right) \left[\xl-1+\redge\left(1-\epsilon \ledge\left(\xl \right)\right)\right], \label{equ:sequation} \\
R_i\left(\xl \right) & = &  \sum_{j \geq 2} \rnode_j \binom{j}{i} \left(\epsilon \ledge\left(\xl \right)\right)^{i} \left(1-\epsilon \ledge\left(\xl \right)\right)^{j-i}, i \geq 2.
\label{equ:tequation}
\end{eqnarray}
Hereby, $L_i\left(\xl \right)$ $\left(R_i(\xl )\right)$ denotes the expected
number of variable (check) nodes of 
degree $i$ at state $\xl$. In the sequel we will refer to these 
equations as {\em density evolution} equations. Rather than considering the evolution of the whole
degree distribution it suffices often to look at some smaller set of parameters. As we have 
discussed, the most important parameter in the decoding process is the number
of degree-one check nodes, denote it by $s\left(\xl\right):=R_1\left(\xl \right)$. 
Further important parameters are the size of the 
residual graph, $v\left(\xl \right):=\sum_{i} L_i\left(\xl \right)$
and the number of check nodes of degree at least two, $t\left(\xl \right):=\sum_{i \geq 2} R_i\left(\xl \right)$.
Let $\nu\left(\xl \right)$, $\sigma\left(\xl \right)$ and $\tau\left(\xl \right)$ denote the respective fractions, 
$v\left(\xl \right)=\lnode(1) \nu\left(\xl \right)$, $s\left(\xl \right)=\lnode(1) \sigma\left(\xl \right)$, $t\left(\xl \right)=\lnode(1) \tau\left(\xl \right)$. 

\begin{figure}[htp]
\begin{center}
\setlength{\unitlength}{1bp}%
\begin{picture}(0,0)
\includegraphics[scale=1]{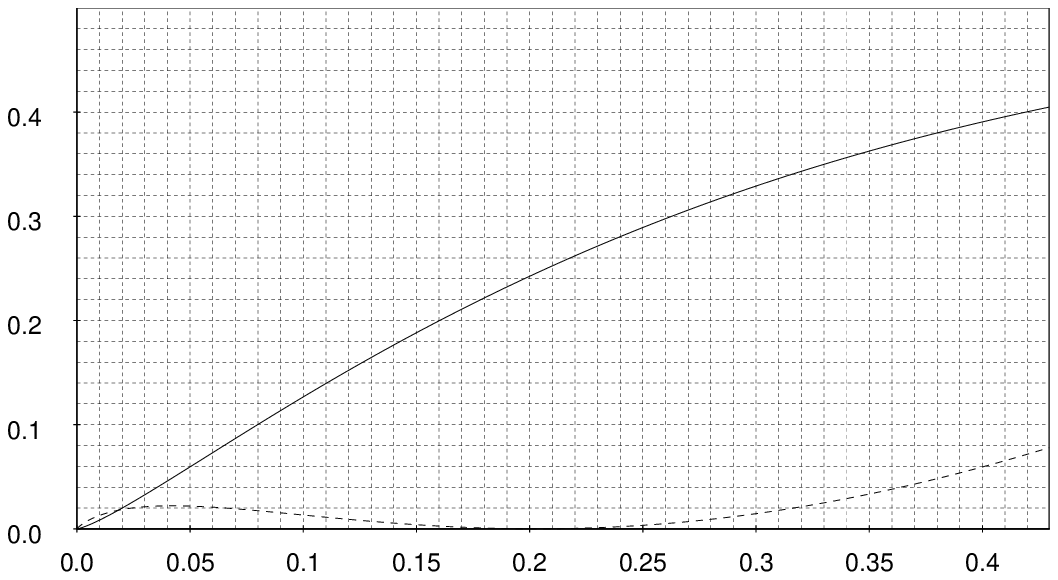}
\end{picture}%
\begin{picture}(320,180)
\put(319, 20){\makebox(0,0){$v$}}
\put(25, 175){\makebox(0,0){$\sigma/\tau$}}
\end{picture}
\end{center}
\caption{\label{fig:evolutionb} The evolution of $\sigma$
and $\tau$ as a function of $\nu$
for the $\eldpc n {x^2} {x^5}$ ensemble and $\epsilon =\epsilon^* \approx 0.4294$.
At $\nu=\nu^* \approx 0.203$, $\sigma(\nu)$ has a minimum and touches the $\nu$-axis.}
\end{figure}

\bexample[Density Evolution of $\eldpc n {x^2} {x^5}$-Ensemble]
Fig.~\ref{fig:evolutionb} depicts the evolution of 
$\sigma$ (dashed line) and $\tau$ (solid line)
as a function of $\nu$ for the ensemble $\eldpc n {x^2} {x^5}$ for the choice
$\epsilon=\epsilon^* \approx 0.4294$. 
Note that for this choice of $\epsilon$
the expected number of check nodes of degree one reaches 
zero at some {\em critical} time of the decoding process.
\EEx
\eexample

The density evolution equations completely specify the asymptotic behavior
of the decoder.
Recall that the decoder stops if the number of degree-one check nodes has reached
zero. If this point is reached before the size of the residual graph has reached zero
a decoding error occurs.
Therefore, if we plot $\sigma(x)$ as a function of $x$ for a given channel parameter
$\epsilon$ we know that the decoder will succeed with high probability if and only
if $\sigma(x)>0$ for all $x \in (0, 1]$.
From equation (\ref{equ:sequation}) we see that
$\sigma(x)>0$ for $x \in (0, 1]$ is equivalent to
\begin{equation} \label{equ:critical}
\redge(1-\epsilon \ledge(x)) > 1-x, \;\;\;\;\;\; \forall x \in (0, 1].
\end{equation}
We can therefore define the {\em threshold} $\epsilon^*(\ledge, \redge)$ as
\[
\epsilon^*(\ledge, \redge) := \sup \{\epsilon: \epsilon \in [0, 1], \redge(1-\epsilon \ledge(x)) > 1-x, \forall x \in (0, 1] \}. 
\]
We say that $x^*$ is a {\em critical} point if $\sigma(x)$ reaches a minimum
at $x=x^*$ and if this minimum is zero, i.e., if  
\[
\redge(1-\epsilon^* \ledge(x^*)) = 1-x^*.
\]

To simplify our matters, we will only discuss ensembles that have a single 
critical point. The extension to several critical points poses no problems in
principle but is technically more cumbersome.
All regular ensembles have this property.
We say that a degree distribution is {\em unconditionally stable} if $x^*>0$, 
i.e., if
the threshold is {\em not} determined by the stability condition.
It is easy to check that this is the
case for all regular ensembles with $\dlmin \geq 3$. Otherwise, i.e., if $x^*=0$ we
say that the ensemble is {\em marginally stable}.
The typical example are cycle code ensembles.
As we will see, the nature of this scaling is drastically different for the two cases.
Finally, we will assume that the degree distributions $\lambda(x)$
and $\rho(x)$ (or just $\lambda(x)$ for Poisson ensembles) are polynomials. 
In this case the density evolution equations have only a finite number
of minima and maxima. This is a purely technical condition to avoid some
pathological cases which are of no practical interest.

\section{Main Results and Discussion}
\label{sec:general}
The following statements apply both to standard ensembles and Poisson 
ensembles.
Generically we will denote such an ensemble by $\eldpc n {\ledge} {\redge}$.
In the Poisson case we can think of $\redge(x)=e^{\frac{x-1}{\bar{r} \int \! \ledge} }$. 

\subsection{Unconditionally Stable Ensembles}

The basic scaling law as given in (\ref{equ:simplscaling}) is stated more precisely in 
the following. 
\blemma[Scaling of Unconditionally Stable Ensembles]
\label{UnconditionalLemma}
Consider transmission over the BEC$(\epsilon)$ using random elements from
an ensemble $\eldpc n {\ledge} {\redge}$
which has a single critical point and is unconditionally stable. 
Let $\epsilon^*=\epsilon^*(\ledge, \redge)$ denote the threshold
and let $\nu^*$ denote the fractional size of the residual graph at the critical point corresponding to the threshold.
Fix $z$ to be $z:=\sqrt{n}(\epsilon^*-\epsilon)$. 
Let $\prob_{\text{b}}(n, \ledge, \redge, \epsilon)$
denote the expected bit erasure probability and let 
$\prob_{\text{B},\gamma}(n, \ledge, \redge, \epsilon)$
denote the expected block erasure probability {\em due to errors of size
at least $\gamma \nu^*$}, where $\gamma \in (0, 1)$.
Then as $n$ tends to infinity,
\begin{align*}
\prob_{\text{B},\gamma}(n, \ledge, \redge, \epsilon) & = Q\left(\frac{z}{\alpha}\right) (1+o_n(1)), \\
\prob_{\text{b}}(n, \ledge, \redge, \epsilon) & = \nu^* Q\left(\frac{z}{\alpha}\right) (1+o_n(1)),
\end{align*}
where $\alpha=\alpha(\ledge, \redge)$
is a constant which depends on the ensemble. 
\elemma
\begin{proof} 
First note that if $\lambda'(0)=0$, i.e., if there are no degree-two variable
nodes, then the block erasure probability is dominated over the whole range of
$\epsilon$ by large error events (when $n$ tends to infinity). 
This means that $\prob_{\text{B},\gamma}$ is equal to 
the ordinary block error probability.

This is no longer true once $\lambda'(0)>0$. If 
$0< \lambda'(0)\rho'(1) <1$ then the
ensemble can be expurgated in order to eliminate small (sublinear weaknesses in
the graph) and the above scaling law will then account for
all errors. If one the other hand no such expurgation is 
done or if $\lambda'(0)\rho'(1) >1$, then besides the contribution to $\prob_{\text{B}}$
stemming from large error events also the contribution stemming from
sublinear-sized weaknesses in the graph will be non-negligible.
The above scaling law only applies to the first contribution.
The bit erasure probability is not affected by these considerations since
the contribution of sublinear-sized stopping sets in the graph vanishes 
as $n$-tends to infinity.
Fortunately, the effect of sublinear-sized stopping sets
is relatively easy to assess
by union bounding techniques. The total erasure probability can
be represented as the sum of these two contributions.
For a more detailed discussion we refer the reader to 
\cite{ZaO02,DRU,RSU02bis}.

Our approach will be to consider first a situation
slightly simplified with respect to the one encountered in iterative decoding.
This will be done in 
Section \ref{sec:parameters} (see Proposition \ref{VarianceEvolutionProp})
and Appendix \ref{se:varapp}. The basic tools needed for the 
proof of this lemma will be introduced in such a simplified context.
It turns out that the main conclusions hold true when the simplifying 
assumptions are removed. This will be shown in Appendix \ref{app:proof}.
\end{proof}

We conjecture that in fact the following refined scaling law is valid.
\bconjecture[Refined Scaling of Unconditionally Stable Ensembles]
\label{RefinedConjecture}
Consider transmission over the BEC$(\epsilon)$ using random elements from
an ensemble $\eldpc n {\ledge} {\redge}$
which has a single critical point and is unconditionally stable.
Let $\epsilon^*=\epsilon^*(\ledge, \redge)$ denote the threshold
and let $\nu^*$ denote the fractional size of the residual graph at the threshold.
Let $\prob_{\text{b}}(n, \ledge, \redge, \epsilon)$
denote the expected bit erasure probability and let
$\prob_{\text{B}, \gamma}(n, \ledge, \redge, \epsilon)$
denote the expected block erasure probability {\em due to errors of size
at least $\gamma \nu^*$}, where $\gamma \in (0, 1)$.
Fix $z$ to be $z:=\sqrt{n} (\epsilon^*-\beta n^{-\frac{2}{3}}-\epsilon)$.
Then as $n$ tends to infinity,
\begin{align*}
\prob_{\text{B}, \gamma}(n, \ledge, \redge, \epsilon)
& = Q\left(\frac{z}{\alpha} \right)\left(1+O(n^{-1/3} \right), \\
\prob_{\text{b}}(n, \ledge, \redge, \epsilon)
& = \nu^* Q\left(\frac{z}{\alpha} \right)\left(1+O(n^{-1/3} \right),
\end{align*}
where $\alpha=\alpha(\ledge, \redge)$ and $\beta=\beta(\ledge, \redge)$
are constants which depend on the ensemble.
\econjecture

This conjecture can be proven in the simplified context mentioned
above (and defined in Section \ref{sec:parameters}).  
This is done in Sec.~\ref{sec:shiftparameters}. At the end of the same
section, we provide some heuristic argument suggesting that the 
simplifying assumptions are in fact irrelevant.

In the remainder of this section we provide  an informal (albeit essentially correct)
justification of the above scaling forms.  
The question of how to compute the scaling parameters will be deferred
to Sections \ref{sec:parameters} (for the variance $\alpha^2$) and
\ref{sec:shiftparameters} (for the shift $\beta$).

Consider the behavior of the individual trajectories of the decoding
process for particular choices of the graph and the channel realization. We will see that
these trajectories closely follow the expected value (given by the density evolution
equations) and that their  standard deviation is of order $\sqrt{n}$.
Consider now the decoding process
and assume that the channel parameter $\epsilon$ is close to $\epsilon^*$.
If $\epsilon=\epsilon^*$ then at the critical point the expected 
number of 
degree-one check nodes is zero. Assume now that we vary $\epsilon$ slightly.
From the density evolution equation (\ref{equ:sequation}) we see that
the expected change in the fraction of degree-one check nodes 
($\sigma = s/n$) at the critical point is 
\begin{equation} 
\label{equ:criticalpoint}
\left.\frac{\partial \sigma}{\partial \epsilon}\right|_{x=x^*; \epsilon=\epsilon^*} =
-\frac{\lnode'(1)}{\lnode(1)} \epsilon^* \ledge(x^*)^2 \redge'(1-\epsilon^* \ledge(x^*)).
\end{equation}
If we vary $\epsilon$ so that $\Delta \epsilon$ is of order $\Theta(1)$, then
we conclude from (\ref{equ:criticalpoint}) that the expected number of degree-one check nodes 
at the critical point is of order $\Theta(n)$. 
Since the standard deviation is of order $\Theta(\sqrt{n})$, 
then with high probability the decoding process will either succeed 
(if $(\epsilon-\epsilon^*)<0$) or die (if $(\epsilon-\epsilon^*)>0$).
The interesting scaling happens if we choose our variation of $\epsilon$
in such a way that
$\Delta \epsilon=z/\sqrt{n}$, where $z$ is a constant.
In this case the expected gap at the critical point scales in the same
way as the standard deviation and one would expect that the probability of
error stays constant. Varying now the constant $z$ will give
rise to the scaling function $f(z)$, cf. equation (\ref{equ:generalfss}). 

\begin{figure}[htp]
\begin{center}
\epsfig{file=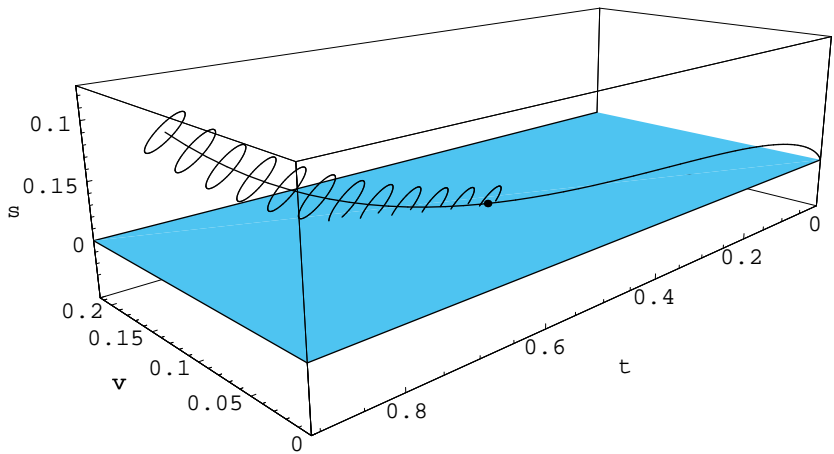, scale=0.9}
\end{center}
\caption{A pictorial representation of density and covariance evolution
for the $\eldpc n {x^2} {x^5}$. Notice that the ellipsoids
corresponding to $(s,t)$ covariances should be regarded as living on a smaller
(by a factor $\sqrt{n}$) scale than the typical trajectory.}
\label{fig:covarianceevolution}
\end{figure}

We will further see that 
the distribution of states at any time before hitting the $s=0$ plane
is Gaussian and that the
evolution of its covariance matrix is governed by a set of
differential equations in the same way as the mean. We will
therefore call these equations the {\em covariance evolution}
equations.
As an example, consider the ensemble $\eldpc n {x^2} {x^5}$
and transmission over the channel BEC$(n, n \epsilon)$. In this
case the residual graph at the start of the decoding process has exactly 
$n \epsilon$
variable nodes and since at each step of the decoding process exactly 
one variable node is pealed off, the size of the residual graph after 
the $\ell$-th decoding
step is exactly $n \epsilon - \ell$ (assuming the decoder has not stopped
prematurely).
As we will discuss in more detail in Section 
\ref{sec:parameters}, it suffices in this case to keep track of the tuple
$(s, t)$ (i.e., we do not need to keep track of the whole degree distribution
of the residual graph). Fig.~\ref{fig:covarianceevolution} shows the
evolution of $(s, t)$ as a function of the size of the residual graph for
the choice $\epsilon=\epsilon^*$. The solid line corresponds to the density 
evolution equation (albeit now in three-dimensional form). 
The dot indicates the critical
point. The ellipsoids represent the covariance matrix. More precisely, they represent
contours of constant probability. 
Note that this picture
is slightly misleading. The ellipsoids
really live on a scale of $\sqrt{n}$ whereas the rest of the graph is scaled
by $n$, i.e., for increasing length the ellipsoids will concentrate more and
more around the expected value.
Those trajectories that hit the $s=0$ plane die. This corresponds
to the part of the ellipsoids that vanish.

One can quantify the probability for the process to hit the
$s=0$ plane as follows. Stop density and covariance evolution
when the number of variables reaches the critical value $v^*$. 
At this point the probability distribution of the state is well approximated
by a Gaussian with a given mean and covariance for $s\ge 0$ (while it is 
obviously 0 for $s<0$). Estimate the survival probability (i.e. the
probability of not hitting the $s=0$ plane at any time) by summing
the Gaussian distribution over $s\ge 0$. Obviously this integral can be 
expressed in terms of a $Q$-function.

We will see that the above description leads indeed to the scaling
behavior as stated in Lemma \ref{UnconditionalLemma}. Where
does the shift in Conjecture \ref{RefinedConjecture} come from?
It is easy to understand that we were a bit optimistic 
(i.e., we underestimated the error probability) in the above calculation:
We correctly excluded from the sum
the part of the Gaussian distribution lying in the $s<0$ half-space --
trajectories contributing to this part must have hit the
$s=0$ plane at some point in the past. On the other hand, we cannot be certain that
trajectories such that $s>0$ when $v$ crosses $v^*$ didn't hit
the $s=0$ plane at some time in the past and bounced back (or will not hit it at some later point). We refer to Section 
\ref{sec:shiftparameters} for an in-depth discussion on how to estimate this effect.

Let us finally recall that the performance over the  BEC$(\epsilon)$  
channel can be easily derived from the results obtained using the model
BEC$(n, n \epsilon)$. One can derive the erasure probability 
for the first case by summing the conditional erasure probability,
where the conditioning is on the number of erasures.
Notice that the number of erasures for the BEC$(\epsilon)$ is asymptotically
Gaussian with mean $n\epsilon$ and standard deviation 
$\sqrt{n \epsilon\overline{\epsilon}}$. Since this standard deviation is of
the same order as the gap to the threshold 
such a convolution gives a non trivial contribution, unlike in the Shannon ensemble
example, cf. Section \ref{sec:introduction}.  It is easy to verify that
this convolution amounts to computing the parameter $\alpha^2$,
cf. Lemma \ref{UnconditionalLemma} as the sum of two contributions:
one due to the channel fluctuations and the other due to covariance evolution.
More precisely we have
\begin{eqnarray}
\alpha_{{\rm BEC}(\epsilon)}^2 =\alpha_{{\rm BEC}(n,n\epsilon)}^2 +\epsilon^*
\overline{\epsilon}^*\, ,\label{equ:channeltranslation}
\end{eqnarray}
where we took $\epsilon =\epsilon^*$ since we are interested in the
region $\epsilon =\epsilon^*+O(n^{-1/2})$ and we can neglect $O(n^{-1/2})$
corrections. Hereafter we shall mostly focus on the BEC$(n, n \epsilon)$
channel. The reader is invited to use the formula 
(\ref{equ:channeltranslation}) for translating the results whenever
necessary.

\subsection{Marginally Stable Ensembles}
As already mentioned, marginally stable ensembles are expected 
to follow a different scaling from the one described in 
Lemma \ref{UnconditionalLemma}.
We will limit our discussion to the simplest case, namely the case of cycle code
ensembles. We conjecture though that the form of the scaling law
is quite general and applies to all marginally stable ensembles.
The cycle Poisson ensemble is slightly easier to handle analytically
than the standard ensemble. We will therefore formulate our results 
mainly for this case.

\blemma[Scaling of Block Probability for Cycle Poisson Ensembles]
\label{lem:blockpoisson}
Consider transmission over BEC$(n, n \epsilon)$ using elements from
$\expeldpc n {\ledge(x)=x} r {s}$. Then
\begin{eqnarray*}
\prob_{\text{B}}(n, \ledge(x)=x, r, s, n \epsilon) =
1- A(s) a n^{-1/6}
\,  f(b\, n^{1/3}(\epsilon-\epsilon^*))\,\left(
	1+O(n^{-1/3})\right) \, ,\nonumber\\
	\label{BlockScalingCycle}
\end{eqnarray*}
where  $a = \bar{r}^{-1/6}$, $b = \bar{r}^{-2/3}$, $A(s) = \exp\left\{\sum_{s'=1}^s \frac{1}{2s'}\right\}$,
and
\begin{eqnarray*}
f(x) = \frac{\sqrt{2\pi} 3^{2/3}}{2}\, e^{-\frac{4x^3}{3}}
p(3^{2/3}x;3/2,-1)\, .
\end{eqnarray*}
Hereby, $p(u;\alpha,\beta)$ is a so called {\em stable density} with representation 
\begin{eqnarray*}
p(u;\alpha,\beta) = \frac{1}{2\pi}\int e^{-itu}\,
\exp\left\{-|t|^{\alpha}e^{-i\frac{\pi}{2}K(\alpha)\beta{\rm sign}(t)}\right\}
\; dt,
\end{eqnarray*}
and $K(\alpha) = 1-\mid 1- \alpha \mid$.
\elemma  
\begin{proof}
In principle one could arrive at the above result by proceeding in the
same fashion as for unconditionally stable ensembles, i.e., one could
employ the tools of density evolution and covariance evolution.

We will however use an entirely different approach. Note that
there is a one-to-one correspondence between elements of 
$\expeldpc n {\ledge(x)=x} r {s=2}$
and random graphs on $n \bar{r}$ nodes
with exactly $n$ edges, see \cite{RSU02}.
If $s=2$, then double edges and cycles of length four are excluded
from the Tanner graph. 
Therefore, each variable node connects two distinct check nodes
and no two variable nodes connect the same pair. If we therefore
identify each variable node (and the two edges that emanate from it)
with one edge in an ordinary graph we get our desired correspondence. 
Further, the decoder
will be successful if and only if this random graph is a {\em forest}, i.e., a
collection of trees. Let $F(l, k)$ denote the number of forests on $l$ labeled
nodes and $k$ components.
Such a forest has $l-k$ edges
and therefore it corresponds to a constellation on $v=l-k$
variable nodes. Since these variable nodes can be ordered
arbitrarily it follows that
there are $v! F(n \bar{r}, n \bar{r}-v)$ constellations
on $v$ variable nodes which do not contain stopping sets. 

It remains to find the total number of constellations on $v$ variable nodes
which are compatible with the expurgation scheme. The desired result will then
follow by diving these two quantities.
Assume $s=0$. Then the total number of constellations on $v$ variable nodes
is equal to $(n \bar{r})^{2 v}$, since for each edge we can choose
one of the $n \bar{r}$ check nodes. Let $n_s(\graph)$ denote the number
of cycles of length $2 s$ in a fixed portion of the bipartite graph $\graph$ of size $v$. It is easy to verify
(and is a well studied problem in random graphs) that 
$\expectation[n_s(\graph)]=\frac{1}{2s}\left(\frac{2 v}{n \bar{r}}\right)^s
(1+O(1/v))$. 
Further it is known that for each fixed $s$ the random variables $(n_1, \cdots n_{s})$
are asymptotically (as $n$ and $v$ tend to infinity with a fixed ratio) independent 
and follow a Poisson distribution, \cite{Bol01}. Finally,
for the Poisson ensemble we have $\epsilon^*=\frac{\bar{r}}{2}$ so that
around the critical value $v = \epsilon^* n = \frac{n \bar{r}}{2}$ and $\frac{2 v}{n \bar{r}}=1$. It follows that 
{\em around the threshold} the total number of constellations which are compatible
with the expurgation scheme behaves like 
\[
T(v \sim n \epsilon^*) = (n \bar{r})^{2 v} e^{-\sum_{s'=1}^{s} \frac{1}{2s'}} (1+O(1/v)) =
(n \bar{r})^{2 v}/A(s) (1+O(1/v)).
\]
From this the block error probability around the threshold follows immediately
once $F(l, k)$ is known, namely, we have
\[
\prob_{\text{B}}(n, \ledge(x)=x, r, s, n \epsilon \sim n \epsilon^*) = 1-A(s)
\frac{(n \epsilon)! F(n \bar{r}, n \bar{r}-n \epsilon)}{(n\bar{r})^{2 n \epsilon}}(1+O(1/n)) \, .
\]
One of the most celebrated formulas in
enumerative combinatorics states that there are $l^{l-2}$ labeled trees on
$l$ nodes, \cite{Wil94}. Unfortunately there does not seem to exist an equally elementary
expression for the number of labeled forests. The situation is aggravated by the fact that 
we are interested in the region where the average number of edges per node
is around one. Exactly around this region the graph goes through a phase
transition and so the behavior of $F(l, k)$ is nontrivial even in the limit of
large sizes.
Fortunately, the asymptotic behavior has been determined by
Britkov \cite{Bri88} and the result has been made accessible (to the
English speaking audience) in the book by Kolchin \cite{Kol99}.
Our result now follows by employing the asymptotic approximation stated in Theorem 1.4.4 
in \cite{Kol99}.\footnote{The reader is warned that there is a slight typo in Theorem 1.4.4 as stated
in \cite{Kol99}.}
\end{proof}

Note, that for the cycle case the maximum likelihood and
the iterative decoder perform {\em identical in terms of block
erasure probability}. This is true since in this case the condition
of no stopping sets is equal to the condition that there are no
cycles which in turns implies that there is no codeword. Note, however,
that this is {\em no longer true once we look at the resulting bit
erasure probability}.

We also note that if we want to get the scaling law for the channel 
BEC$(\epsilon)$ we need to 
convolve the above curves with the Binomial with mean $n \epsilon$. 
However, on the scale $\epsilon^*-\epsilon = O(n^{-1/3})$, the
effect of the channel fluctuations vanishes in the large blocklength limit.
The leading correction to the scaling law (\ref{BlockScalingCycle})
coming from the channel consists in the substitution
\begin{eqnarray}
f(x)\to f(x) + \frac{\epsilon^*(1-\epsilon^*)}{(1-r)^{4/3}} \, 
f''(x) \, n^{-1/3}\, +O(n^{-1/2})\, .
\end{eqnarray}

\begin{figure}[htp]
\begin{center}
\setlength{\unitlength}{0.4bp}%
\begin{picture}(0,0)
\includegraphics[scale=0.4]{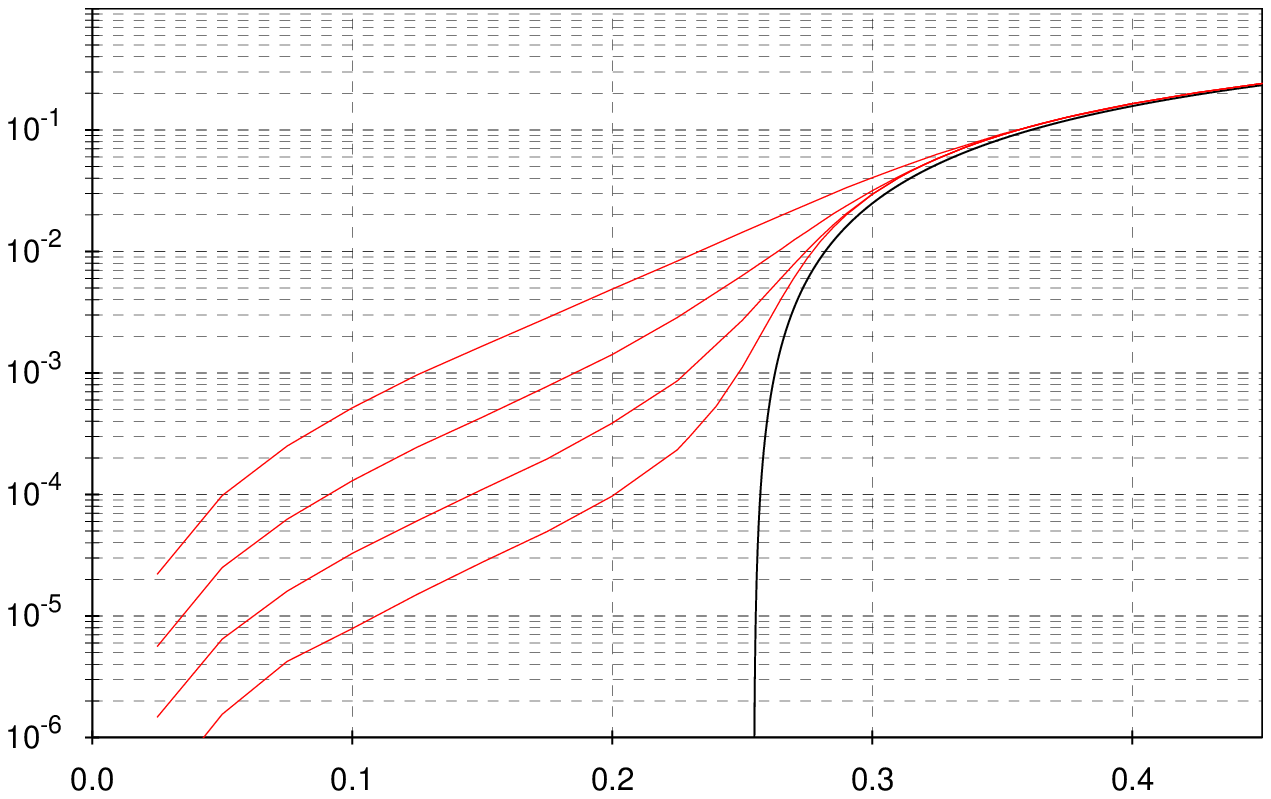}
\end{picture}%
\begin{picture}(367.5,231)
\put(13,223){\makebox(0,0){$\prob_{\text{b}}$}}
\put(366,8){\makebox(0,0)[r]{$\epsilon$}}
\end{picture}
\hspace{0.2in}
\setlength{\unitlength}{0.4bp}%
\begin{picture}(0,0)
\includegraphics[scale=0.4]{PBPoiss0.5.ps}
\end{picture}%
\begin{picture}(367.5,231)
\put(13,223){\makebox(0,0){$\prob_{\text{B}}$}}
\put(366,8){\makebox(0,0)[r]{$\epsilon$}}
\end{picture}
\end{center}
\caption{\label{fig:cyclecodes}
The bit and block erasure probability for $\expeldpc n {\ledge(x)=x} {r=\frac{1}{2}} {s=1}$
for $n=2^i$, $i=8, 10, 12, 14$.
As can be seen from the picture, the block erasure curves
actually converge to a limiting (non-zero) curve over the whole range of $\epsilon$,
whereas the bit erasure curves decrease to zero below the threshold for increasing block lengths.
Also shown are the result of using the scaling laws for the block erasure probability
as stated in Lemma \ref{lem:blockpoisson}.} 
\end{figure}

The following lemma characterizes the corresponding limiting block erasure probability curve.
\blemma[Asymptotic Block Erasure Probability Curve]
Consider transmission over BEC$(n, n \epsilon)$ or BEC$(\epsilon)$ using random elements from
$\expeldpc n {\ledge(x)=x} r s$.
Then
\[
\lim_{n \rightarrow \infty} \prob_{\text{B}}(n, \ledge(x)=x, r, s, n \epsilon) = 1- 
\sqrt{1-\frac{\epsilon}{\epsilon^*}}
\exp\left\{\sum_{s'=1}^s\frac{\left(\frac{\epsilon}{\epsilon^*}\right)^{s'}}{2 s'} \right\}.
\]
\elemma
The corresponding asymptotic bit erasure probability curve under iterative
decoding can be obtained through a standard density evolution analysis and
it is given in parametric form by
\[
\left(
\frac{x}{\ledge(1-\redge(1-x))}, 
\frac{x \lnode(1-\redge(1-x))}{\ledge(1-\redge(1-x))}
\right),
\]
where $x \in (x^*, 1]$ and $x^*$ is the solution to the equation $\epsilon^* \ledge(1-\redge(1-x))=x$.
Figure \ref{fig:cyclecodes} shows the resulting bit and block erasure curves for
$\expeldpc n {\ledge(x)=x} {r=\frac{1}{2}} {s=1}$.

\begin{figure}[htp]
\begin{center}
\setlength{\unitlength}{0.6bp}%
\begin{picture}(0,0)
\includegraphics[scale=0.6]{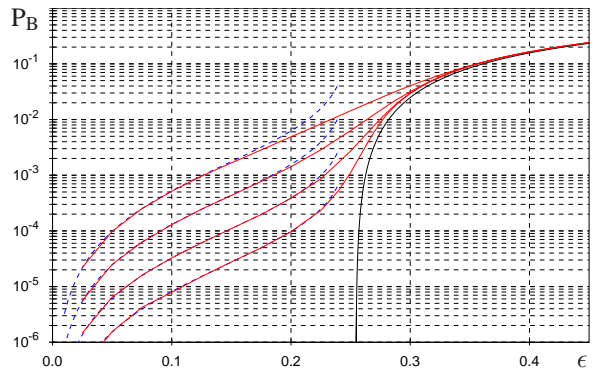}
\end{picture}%
\begin{picture}(367.5,231)
\put(13,223){\makebox(0,0){$\prob_{\text{B}}$}}
\put(366,8){\makebox(0,0)[r]{$\epsilon$}}
\end{picture}
\end{center}
\caption{\label{fig:cyclecodesfloor}
Comparison of the exact bit erasure curves (solid line) with the analytic expression
given in (\ref{equ:errorfloor}) (dashed lines) for $n=2^i$, $i=8, 10, 12, 14$ and $\epsilon
< \epsilon^*$.}
\end{figure}

Cycle codes can not be expurgated up to some linear fraction of the block length
since the number of stopping sets of size $s_1, \cdots s_k$ are jointly Poisson
and have mean equal to $(2/\bar{r})^{s_i}/(2 s_i)$, respectively. Below the 
threshold $\epsilon^* = \bar{r}/2$, the bit erasure probability scales 
as $1/n$. Expurgation changes uniquely the coefficient of this scaling.
A simple calculation yields
\begin{eqnarray} \label{equ:errorfloor}
\prob_{\text{b}}(n, \ledge(x)=x, r, s, n \epsilon) = \frac{1}{2n}
L_s\left(\frac{2\epsilon}{\bar{r}} \right) (1+ O(1/n))\, ,\label{equ:cyclicerrorfloor}
\end{eqnarray}
where we defined the function 
\[
L_s(x) := \sum_{s'=s+1}^{\infty} \frac{x^{s'}}{s'} = -\log(1-x)-
\sum_{s'=1}^{s} \frac{x^{s'}}{s'}\, .
\]
As shown in Fig. \ref{fig:cyclecodesfloor}, this formula provides a good 
approximation to the bit error probability {\em away} from the critical 
region. Notice in fact that  the coefficient of the $1/n$ term in Eq. 
(\ref{equ:cyclicerrorfloor}) diverges as $\epsilon\to\epsilon^*$.

\section{Computation of the Variance Parameter}
\label{sec:parameters}
In the previous section we saw that the basic scaling
law, cf. Lemma \ref{UnconditionalLemma}, only depends on the variance 
$\alpha^2$.
In this section we will work out in detail the calculation of
this parameter. In Section \ref{sec:shiftparameters} we will present the method to be used for computing 
$\beta$ which is needed for the conjectured refined form of the scaling law.

Although conceptually it is straightforward to write down
the equations for the general irregular case, the actual computations
are quite cumbersome. We will therefore proceed as follows. In 
Section~\ref{sec:GeneralCovarianceEvolution} we discuss the covariance evolution
equations in an abstract setting. 
These are applied to particular regular LDPC ensembles in 
Section~\ref{RegularParametersSec}.
%
%
\subsection{General Covariance Evolution}
\label{sec:GeneralCovarianceEvolution}

We regard iterative decoding as a Markov process in a 
finite dimensional space. The examples in the next two subsections will make 
clear how this framework can be adapted to particular code ensembles.

Consider a family of Markov chains $X_{n,0},X_{n,1},\dots,X_{n,t},\dots$ parametrized
by $n\in{\mathbb N}$ and taking values in ${\mathbb Z}^{d+1}$. 
For iterative decoding applications, 
$n$ will represent the blocklength.
We drop the subscript $n$ hereafter. Let the transition probability be
\begin{eqnarray}
\prob(X_{t+1}=x'|X_t=x) = W(x'-x|x)\, ,
\end{eqnarray}
and the initial condition be a single non-random state $X_0 = 
x_0\in{\mathbb Z}^{d+1}$. In iterative decoding the initial condition
is actually a distribution over states. This case is easy to treat by 
first conditioning on the initial state, and then convolving with the
initial distribution. 
We will denote the $d+1$ coordinates of the state $x$ as 
\begin{eqnarray}
(x^{(0)},x^{(1)},\dots, x^{(d)}) = x\in{\mathbb Z}^{d+1}\,.
\end{eqnarray}
We denote the corresponding random variable by
$(X^{(0)},X^{(1)},\dots,X^{(d)})$.

In the following we shall always be 
interested in times $t< \kappa_0\, n$ for a positive constant 
$\kappa_0$ (we reserve the symbols $\kappa_1,\kappa_2,\dots$
for numerical constants which we assume not to depend upon $n$).
We shall moreover assume the following regularity properties of the 
Markov chain:
\begin{enumerate}
\item The chain makes finite jumps. In other words, there exists a 
$\kappa_1>0$ such that $|X^{(i)}_{t+1}-X^{(i)}_{t}|<\kappa_1$ almost surely.
\item The transition probabilities have a smooth $n\to\infty$ limit. 
In practice there exist functions 
$\Wh: {\mathbb Z}^{d+1}\times {\mathbb R}^{d+1} \to  {\mathbb R}_+$
and a positive constant $\kappa_2$ such that
\begin{eqnarray}
|W(\Delta|x)-\Wh(\Delta| x/n)|<\kappa_2/n\, .
\end{eqnarray}
Clearly, we have $\sum_{\Delta} \Wh(\Delta| x/n)=1$. 
We shall moreover assume $\Wh(\Delta| z)$ to be 
$C^2({\mathbb R}^{d+1})$ with respect to its second argument and to have 
bounded first and second derivatives.
\item The process has a finite range on the $n$ scale. In practice,
there exists $\kappa_3>0$ such that $|X^{(i)}_t|<\kappa_3\, n$ almost surely.
\end{enumerate}

Under these hypothesis the distribution of $X_t$ is well described
by a Gaussian whose mean and variance can be obtained by solving 
some ordinary differential equations. In order to state this 
fact in a more precise fashion, we need some additional notation.
We denote by $\oX_t \equiv \expectation[ X_t]$ the average of $X_t$ and
$D^{(ij)}_t \equiv \expectation[ X^{(i)}_t;X^{(j)}_t]
\equiv  \expectation[ X^{(i)}_t X^{(j)}_t] -  \expectation[ X^{(i)}_t]
\expectation[ X^{(j)}_t]$ its covariance.
We need furthermore the first two moments of the transition rates
$W(\Delta|x)$:
\begin{eqnarray}
f^{(i)}(x) &\equiv& \sum_{\Delta} \Delta_i \, W(\Delta|x)\, ,\label{DriftDefinition}\\
f^{(ij)}(x) & \equiv & \sum_{\Delta} \Delta_i\Delta_j \, W(\Delta|x)-
f^{(i)}(x) f^{(j)}(x) \, ,\label{DiffusionDefinition}
\end{eqnarray}
with $i,j\in \{0,\dots,d\}$. We shall call $\fh^{(i)}(z)$, $\fh^{(ij)}(z)$
the analogous quantities for the limiting rates $\Wh(\Delta|z)$.

Finally, let $\oz(\tau)\in {\mathbb R}^{d+1}$ and 
$\delta^{(ij)}(\tau)\in {\mathbb R}$, for $\tau\in {\mathbb R}_+$
and $i,j\in\{0,\dots, d\}$, denote the solution of
\begin{eqnarray}
\frac{d\oz^{(i)}}{d\tau}(\tau) & = & \fh^{(i)}(\oz(\tau))\, ,
\label{GeneralDensityEvolution}\\
\frac{d\delta^{(ij)}}{d\tau}(\tau) & = & \fh^{(ij)}(\oz(\tau)) + 
\sum_{k=0}^d\left[\delta^{(ik)}(\tau)
\left.\frac{\partial \fh^{(j)}}{\partial z^{(k)}}\right|_{\oz(\tau)} 
+ \left.\frac{\partial \fh^{(i)}}{\partial z^{(k)}}\right|_{\oz(\tau)} 
\delta^{(kj)}(\tau)\right]\, .\label{GeneralCovarianceEvolution}
\end{eqnarray}
with initial conditions $\oz(0) = x_0/n$ and $\delta^{(ij)}(0) = 0$.
\begin{proposition}
\label{VarianceEvolutionProp}
Under the conditions stated above the following results hold
(here we use the symbols $\Omega_0, \Omega_1, \dots$, for 
constants (independent of $n$) which we prove to exist):
\begin{enumerate}
\item[I.] $X_t$ concentrates on the $n$ scale. In formulae, there exist 
$\Omega_0>0$, such that
\begin{eqnarray}
\prob\{|X_t^{(i)}-\oX_t^{(i)}|\ge \rho\}\le 2\, e^{-\frac{\rho^2}{2\Omega_0 t}}.
\label{Concentration}
\end{eqnarray}
\item[II.] The average and covariance of $X_t$ are accurately tracked by
$\oz(\tau)$ and $\delta^{(ij)}(\tau)$. More precisely, there exist
constants $\Omega_1,\Omega_2>0$, such that
\begin{eqnarray}
\left|\frac{1}{n}\oX^{(i)}_t-\oz^{(i)}(t/n)\right| & \le& \frac{\Omega_1}
{n}\, ,
\label{FirstMomentEstimate}\\
\left|\frac{1}{n}D^{(ij)}_t-\delta^{(ij)}(t/n)\right| & \le &
\frac{\Omega_2}{\sqrt{n}}. 
\label{SecondMomentEstimate}
\end{eqnarray}
\item[III.] The variable $(X_t-\oX_t)/\sqrt{n}$ converges weakly
to a $(d+1)$-dimensional Gaussian with variance $\delta^{(ij)}(t/n)$. More
precisely, define the logarithmic moment generating function
\begin{eqnarray}
\Lambda_t(\lambda) \equiv \log \expectation\, \exp\left[\frac{1}{\sqrt{n}}
\, \, \lambda\cdot (X_t-\oX_t)\right],
\label{GeneratingFunctionDefinition}
\end{eqnarray}
for $\lambda\in{\mathbb R}^{d+1}$. Then there exist a function 
$\lambda\mapsto \Omega_4(\lambda)\in{\mathbb R}_+$, such that
\begin{eqnarray}
\left|\Lambda_t(\lambda) - \frac{1}{2}\sum_{ij}\delta^{(ij)}(t/n)
\lambda_i\lambda_j\right|\le \frac{\Omega_4(\lambda)}{\sqrt{n}}\, .
\label{GeneratingFunctionEstimate}
\end{eqnarray}
\end{enumerate}
\end{proposition}
The proof is quite straightforward and will be 
outlined in App.~\ref{se:varapp}. Here we limit ourselves to a few comments. 

Notice that the statements collected in the above proposition are not all 
independent. Equation (\ref{SecondMomentEstimate}), may for instance
be regarded as a consequence of Eq. (\ref{GeneratingFunctionEstimate}).
The various results are presented in order of increasing sharpness.
Also, not all of the assumptions in the points 1-3 are needed to
proof each of the statements in the proposition. For instance, the 
concentration result is an easy consequence of the Hoeffding-Azuma inequality
and requires the hypotheses 1 (uniformly bounded jumps), 
3 (scaling of time with $n$) plus some Lipschitz property of
the drift coefficients $f^{(i)}(x)$, cf. Eq. (\ref{DriftDefinition}).
This point is further discussed in App.~\ref{se:varapp}.
The limitation to a deterministic initial condition is easily 
removed. In iterative decoding applications the initial condition is
a Gaussian distribution with standard deviation of order 
$\sqrt{n}$. Convolution with such a distribution amounts to 
integrating equation (\ref{GeneralCovarianceEvolution}) and taking
as initial condition the initial covariance.  
Finally, the situation investigated here can be regarded as a 
discrete analogous of the Friedlin-Wentzell theory of random perturbations
of dynamical systems \cite{FrW84}.

In the following section we shall apply the above analysis to
two LDPC ensembles: the standard regular ensemble 
$\eldpc {n} {x^{\dl-1}} {x^{\dr-1}}$, and the regular Poisson
ensemble $\eldpc {n} {x^{\dl-1}}{r}$. The general strategy is the 
following: $(i)$ Determine a sufficient statistics for the decoding process.
For a  general $\eldpc {n} {\lambda} {\rho}$ ensemble, a sufficient
statistics is provided by the degree distributions at variable and check 
nodes in the residual graph. As we will see, a more compact representation
is available for the two special cases mentioned above.
$(ii)$ Write
the transition probability for iterative decoding and compute
the drift and diffusion coefficients, cf. Eqs. (\ref{DriftDefinition}), 
(\ref{DiffusionDefinition}).
$(iii)$ Determine the initial condition, namely the average state, and its
variance before the decoding process has been started.  
$(iv)$ Integrate the density evolution
and covariance evolution equation, cf. Eq. (\ref{GeneralDensityEvolution})
and (\ref{GeneralCovarianceEvolution}) up to the critical point.
The parameter $\alpha$ in Lemma \ref{UnconditionalLemma} is finally 
given (up to a rescaling) by the standard deviation of the number of degree one check
nodes $s$ at the critical point. 
More precisely:
\begin{eqnarray}
\alpha = 
\sqrt{\delta_{\sigma \sigma}} \left(\frac{\partial\sigma}{\partial \epsilon}\right)^{-1}\, ,\label{equ:alphaformula}
\end{eqnarray}
both factors being evaluated at the critical point.
%
%
\subsection{Regular Ensembles}
\label{RegularParametersSec}
We will now show the explicit computations that need to be done in order to accomplish the
program outlined in the previous section for the case of regular standard and Poisson ensembles.

There are some significant simplifications that arise in this case. Note that
the triple $(v, s, t)$ constitutes a sufficient statistics, i.e., it suffices to keep
track of the number of variable nodes (all of which have degree $\dl$ since by assumption the
graph is regular), the number of degree-one check nodes and the number of check nodes of
degree two or higher. This can be seen as follows. We claim that all constellations of
``type'' $(v, s, t)$ have uniform probability. To see this let $\tilde{\graph}_1$ 
and $\tilde{\graph}_2$ be two residual graphs of type $(v, s, t)$. Assume that $\tilde{\graph}_1$
is the result of applying the iterative decoder to the graph $\graph_1$
with a particular channel realization and a particular sequence of choices of the iterative decoder.
It is then easy to see that there exists a graph $\graph_2$ which differs from $\graph_1$
only on the residual part (where it coincides with $\tilde{\graph}_2$) 
but agrees with it otherwise. By definition of the ensemble,
$\graph_1$ and $\graph_2$ have equal probability and if the iterative decoder is applied to
$\graph_2$ with the same channel realization and sequence of random choices we get $\tilde{\graph}_2$.
This shows that $\tilde{\graph}_1$ and $\tilde{\graph}_2$ (and therefore any residual graph
which is compatible with the degree distribution) have equal probability.
It follows that, given $(v, s, t)$, the distribution of $\graph$ is determined so
that $(v, s, t)$ indeed constitutes a state.

Let us now determine the degree distribution of a ``typical'' element $\graph$
of type $(v, s, t)$, since this knowledge will be required in the sequel.
For the standard ensemble
define the generator polynomial $p(z):= (1+z)^{\dr}-\dr z -1$ which counts the number
of connections into a check node of degree two or higher. For the Poisson ensemble the equivalent function
is $p(z):=e^z-z-1$. Define $a(z):= z \frac{p'(z)}{p(z)}$.
The total number of constellations on $t$ check nodes of degree at least two with  $v \dl-s$ edges is easily seen to be 
$\text{coef}\left\{ p(x)^t, x^{v \dl -s} \right\}$. Let $t_i$, $i \geq 2$, denote the number of check nodes of degree $i$.
Then the total number of constellations which are compatible with the desired type can be written as
\begin{eqnarray*}
\sum_{t_2, t_3, \cdots: \sum_{i \geq 2} t_i=t; \sum_{i \geq 2} i t_i= v \dl- s} \binom{t}{t_2, t_3, \cdots} \left(\prod_{i \geq 2} p_i^{t_i} \right).
\end{eqnarray*}
Since all constellations have equal probability a ``typical'' constellation will have the type
which ``dominates'' the above sum. Some calculus reveals that this dominating type has the form
\begin{equation} 
\label{equ:degreedistribution}
\tau_i= \frac{p_i z^i}{p(z)} \tau\, ,\;\;\;\;\;\; i \geq 2,
\end{equation}
where $\tau_i$, $i \geq 2$, denotes the fraction of check nodes of degree $i$
and where $a(z)=\frac{\nu \dl-\sigma}{\tau}$.

We will see shortly that for the Poisson case it suffices to consider ensembles
of rate zero since the scaling parameters for the general case can be easily 
connected to this case. Therefore in the next theorem we can assume without loss of
generality that the rate is zero for Poisson ensembles.
 
\label{StandardRegularParametersSec}
\blemma[Drift, Variance and Partial Derivatives for Regular Ensembles] 
\label{StandardRegularParametersLemma}
Consider regular standard ensembles $\eldpc n {x^{\dl-1}} {x^{\dr-1}}$
or regular Poisson ensembles $\eldpc n {x^{\dl-1}} {r=0}$.
Define 
\begin{eqnarray*}
p(z) & = &
\begin{cases}
(1+z)^{\dr}-1-\dr z, & \text{standard ensemble}, \\
e^z-1-z, & \text{Poisson ensemble},
\end{cases} 
\end{eqnarray*}
and let $a(z):= z \frac{p'(z)}{p(z)}$. Let $\xleft$ denote the right-to-{\em left} erasure probability
and let $\xright:=\epsilon \ledge(\xleft)$. Then
along the density evolution path parametrized by $\xleft$ we have
\begin{align*}
\hat{f}^{(\tau)} & = - (\dl-1) \frac{2 \tau_2}{\nu \dl}, & 
\hat{f}^{(\sigma)} & = -1-(\dl-1) \frac{\sigma}{\nu \dl}- \hat{f}^{(\tau)}, \\
\hat{f}^{(\tau \tau)} & = -\hat{f}^{(\tau)} \left( 1+\frac{\hat{f}^{(\tau)}}{\dl-1} \right), &
\hat{f}^{(\sigma \tau)} & = \hat{f}^{(\tau)} \left(1-\frac{\hat{f}^{(\sigma)}+ 1}{\dl-1} \right)\, ,\\
\frac{\partial \hat{f}^{(\tau)}}{\partial \sigma} & =  \frac{2 (\dl-1)}{\nu \dl}
\frac{p_2 z (2 - a(z))}{a'(z) p(z)}, & 
\frac{\partial \hat{f}^{(\sigma)}}{\partial \nu} & = \frac{\dl-1}{\nu \dl} \frac{\sigma}{\nu} - \frac{\partial \hat{f}^{(\tau)}}{\partial \nu}\\
\frac{\partial \hat{f}^{(\sigma)}}{\partial \tau} & =  -\frac{\partial \hat{f}^{(\tau)}}{\partial \tau} , & 
\frac{\partial \hat{f}^{(\sigma)}}{\partial \sigma} & =  - \frac{\dl-1}{\nu \dl}-\frac{\partial \hat{f}^{(\tau)}}{\partial \sigma} , \\
\end{align*}
\vspace{-1.4cm}
\begin{eqnarray*}
\hat{f}^{(\sigma \sigma)} & = & -\frac{\left(\hat{f}^{(\tau)}\right)^2}{\dl-1}- (\dl-1) \left(\frac{\sigma}{\nu \dl}-1 \right) \frac{\sigma}{\nu \dl} - 
\hat{f}^{(\tau)} \left(1 + 2 \frac{\sigma}{\nu \dl} \right), \\
\frac{\partial \hat{f}^{(\tau)}}{\partial \nu} & = & -\frac{2 (\dl-1)}{\nu \dl} \left( - \frac{\tau_2}{\nu} + \frac{p_2 z \dl (2-a(z)) }{a'(z)p(z)}   \right) \\
\frac{\partial \hat{f}^{(\tau)}}{\partial \tau} & = & -\frac{2 (\dl-1)}{\nu \dl} \left(\frac{\tau_2}{\tau}-
\frac{p_2 z a(z)(2 - a(z))}{a'(z) p(z)} \right), \\
\end{eqnarray*}
where for the standard regular ensemble $z=\frac{\epsilon \ledge(\xleft)}{1-\epsilon \ledge(\xleft)}$ whereas for
the Poisson regular ensemble $z = \frac{\epsilon \ledge(\xleft)}{\int \ledge}$.
\elemma
\begin{proof}
Let $\sigma$ denote the fraction of degree-one check nodes, $\tau_i$,
$i \geq 2$, the fraction of degree-$i$ check nodes and $\nu$ denote the fraction
of residual variable nodes. Since the total edge
count on the left and right must match up we have $\sigma+\sum_{i=2} i \tau_i= \nu \dl$.
A random edge therefore has probability $q_1:=\frac{\sigma}{\nu \dl}$ of being connected to
a degree-one check node and probability $q_i:=\frac{i \tau_i}{\nu \dl}$ of being connected to a degree-$i$
check node, $i \geq 2$. For large $n$, the joint probability distribution of all $\dl$
edges emanating from a variable node converges to the product distribution. It follows
that (in this large blocklength limit) the probability distribution (for a randomly chosen
variable node) of having $u_1$ connections
into degree-one check nodes and $u_2$ connections into degree-two check nodes is given by
\[
\tilde{w}(u_1, u_2):= \binom{\dl}{u_1, u_2, \dl-u_1-u_2} q_1^{u_1} q_2^{u_2} (1-q_1-q_2)^{\dl-u_1-u_2}.
\] 
In the iterative decoding process variables are not picked at random though.
A variable node is picked with a probability which is proportional to $u_1$.
Therefore, the induced probability distribution under iterative decoding is 
\begin{eqnarray}
w(u_1, u_2) = \frac{\tilde{w}(u_1, u_2) u_1}{\sum_{u_1', u_2'} \tilde{w}(u_1', u_2') u_1'} = \tilde{w}(u_1, u_2) \frac{u_1}{\dl q_1}\, ,
\label{eq:ContinuumTransition}
\end{eqnarray}
Note that the generating function of $w(u, v)$ 
has the compact description
\begin{eqnarray*}
W(x, y) & := & \sum_{u_1, u_2} w(u_1, u_2) x^{u_1} y^{u_2} = x (x q_1 + y q_2 +(1-q_1-q_2))^{\dl-1}.
\end{eqnarray*}
In terms of $W(x, y)$ we have
\begin{eqnarray*}
\hat{f}^{(\tau)} & = & - \sum_{u_1, u_2} w(u_1, u_2) u_2 =
- \left. \frac{\partial W(x, y)}{\partial y}\right|_{x=y=1} \\
& = & -(\dl-1) q_2 = - (\dl-1) \frac{2 \tau_2}{\nu \dl}, \\
\hat{f}^{(\sigma)} & = & - \sum_{u_1, u_2} w(u_1, u_2) (u_1-u_2) = 
- \left. \frac{\partial W(x, y)}{\partial x}\right|_{x=y=1}-\hat{f}^{(\tau)} \\
& = & -1-(\dl-1) q_1 - \hat{f}^{(\tau)} = - 1- (\dl-1) \frac{\sigma}{\nu \dl} -\hat{f}^{(\tau)}, \\
\hat{f}^{(\tau \tau)} & = & \sum_{u_1, u_2} w(u_1, u_2) u_2^2 -\left(\hat{f}^{(\tau)}\right)^2 =
\left. \frac{\partial^2 W(x, y)}{\partial y^2}\right|_{x=y=1}-\hat{f}^{(\tau)}-\left(\hat{f}^{(\tau)}\right)^2  \\
& = & (\dl-1) (\dl-2) q_2^2-\hat{f}^{(\tau)}-\left(\hat{f}^{(\tau)}\right)^2= -\hat{f}^{(\tau)} \left(1+ \frac{\hat{f}^{(\tau)}}{\dl-1} \right), \\
\hat{f}^{(\sigma \tau)} & = & \sum_{u, v} w(u_1, u_2) (u_1-u_2) u_2
-\hat{f}^{(\sigma)} \hat{f}^{(\tau)} =
\left. \frac{\partial^2 W(x, y)}{\partial x y}\right|_{x=y=1}-\hat{f}^{(\tau \tau)}- \left(\hat{f}^{(\tau)}\right)^2-
\hat{f}^{(\sigma)} \hat{f}^{(\tau)} \\
& = & -\hat{f}^{(\tau)}(1+(\dl-2) q_1)- \hat{f}^{(\tau \tau)}-\left(\hat{f}^{(\tau)}\right)^2-\hat{f}^{(\sigma)} \hat{f}^{(\tau)} = 
\hat{f}^{(\tau)} \left(1-\frac{\hat{f}^{(\sigma)}+ 1}{\dl-1} \right)  \\
\hat{f}^{(\sigma \sigma)} & = & \sum_{u_1, u_2} w(u_1, u_2) (u_1-u_2)^2 -\left(\hat{f}^{(\sigma)}\right)^2 \\ 
& = & \left. \frac{\partial^2 W(x, y)}{\partial x^2}\right|_{x=y=1}
-\hat{f}^{(\sigma \tau)}-\hat{f}^{(\sigma)} \hat{f}^{(\tau)}-\hat{f}^{(\tau \tau)}-\left(\hat{f}^{(\tau)}\right)^2-\hat{f}^{(\sigma)}-\hat{f}^{(\tau)}- \left(\hat{f}^{(\sigma)}\right)^2 \\
& = & (\dl-1) q_1 \left(2+(\dl-2) q_1 \right)-\hat{f}^{(\sigma \tau)}-\hat{f}^{(\sigma)} \hat{f}^{(\tau)}-\hat{f}^{(\tau \tau)}-
\left(\hat{f}^{(\tau)}\right)^2-\hat{f}^{(\sigma)}-\hat{f}^{(\tau)}- \left(\hat{f}^{(\sigma)}\right)^2 \\
& = & -\frac{\left(\hat{f}^{(\tau)}\right)^2}{\dl-1}- (\dl-1) \left(\frac{\sigma}{\nu \dl}-1 \right) 
\frac{\sigma}{\nu \dl} - \hat{f}^{(\tau)} \left(1 + 2 \frac{\sigma}{\nu \dl} \right) .
\end{eqnarray*}

Next we need to determine the partial derivatives. From equation (\ref{equ:degreedistribution}) 
for $i=2$ we have
\begin{eqnarray*}
\frac{\partial \hat{f}^{(\tau)}}{\partial \tau} & = & 
- \frac{2 (\dl-1)}{\nu \dl} \frac{\partial \tau_2}{\partial \tau} 
 =  - \frac{2 (\dl-1)}{\nu \dl} \left( \frac{\tau_2}{\tau} + 
\tau \frac{\partial p_2 \frac{z^2}{p(z)}}{\partial z}
\frac{\partial z}{\partial \tau}  \right)  \\
& = & -\frac{2 (\dl-1)}{\nu \dl} \left(\frac{\tau_2}{
	\tau}-
	\frac{p_2 z a(z)(2 - a(z))}{a'(z) p(z)} \right).
\end{eqnarray*}
The remaining derivatives follow in the same way and we skip the details.
Now note that along the typical decoding trajectory  
all quantities required to compute the above expressions are given
by the density evolution equations (\ref{equ:sequation}) and (\ref{equ:tequation}).

It remains to establish the link between $z$ and $\xleft$. We start with
standard ensembles. From the density evolution equation (\ref{equ:tequation})
\begin{eqnarray*}
\tau_2 & = & \frac{\binom{\dr}{2} \xright^2 (1-\xright)^{\dr-2}}{
\sum_{i \geq 2} \binom{\dr}{i} \xright^i (1-\xright)^{\dr-i}} \tau 
= \frac{\binom{\dr}{2} \left(\frac{\xright}{1-\xright} \right)^2 }{
  \sum_{i \geq 2} \binom{\dr}{i} \left(\frac{\xright}{1-\xright} \right)^i} \tau 
  = \frac{p_2 \left(\frac{\xright}{1-\xright} \right)^2}{p\left(\frac{\xright}{1-\xright} \right)} \tau.
\end{eqnarray*}
Comparing this to equation (\ref{equ:degreedistribution})
it follows that $z = \frac{\xright}{1-\xright}=\frac{\epsilon \ledge(\xleft)}{1-\epsilon \ledge(\xleft)}$.

Recall that in the Poisson case we can assume that $r=0$, so that
$\redge(x)=R(x)=e^\frac{x-1}{\int \ledge}$. Again from (\ref{equ:tequation}) 
\begin{eqnarray*}
\tau_2 & = & \frac{\xright^2}{2 (\int \lambda)^2} R(1-\xright) 
= \frac{p_2 \left( \frac{\xright}{\int \ledge} \right)^2}{p\left( \frac{\xright}{\int \ledge}\right)} \tau,
\end{eqnarray*}
from which it follows that for the Poisson case
\[
z = \frac{\epsilon \ledge (\xleft)}{\int \ledge}.
\]
\end{proof}

\begin{figure}[htp]
\begin{center}
\setlength{\unitlength}{1bp}%
\begin{picture}(0,0)
\includegraphics[scale=1.0]{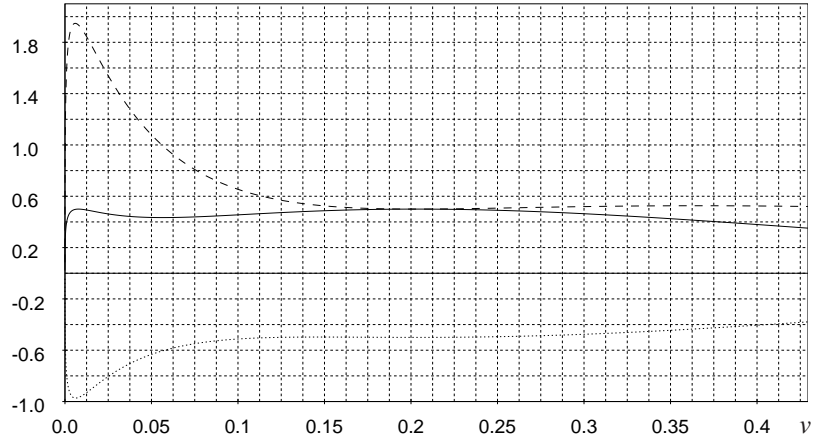}
\end{picture}%
\begin{picture}(320,180)
\put(319, 20){\makebox(0,0){$v$}}
\end{picture}
\end{center}
\caption{\label{fig:initialvariance} The evolution of 
$\hat{f}^{(\sigma \sigma)}$ (dashed line), $\hat{f}^{(\sigma \tau)}$ (dotted line)
and $\hat{f}^{(\tau \tau)}$ (solid line) along the critical trajectory for 
the $\eldpc n {x^{2}} {x^{5}}$ ensemble.}
\end{figure}

Figure \ref{fig:initialvariance} depicts $\hat{f}^{(\sigma \sigma)}$, $\hat{f}^{(\sigma \tau)}$ and 
$\hat{f}^{(\tau \tau)}$ as a function of $\nu$ along the critical trajectory 
(i.e., for the choice $\epsilon=\epsilon^* \approx 0.4294$) for the 
$\eldpc n {x^{2}} {x^{5}}$ ensemble.

The last piece of information required to apply the 
strategy outlined in the previous subsection, consists in determining the 
initial condition for the density and covariance evolution.
This is provided by the following lemmas, whose proof are fairly routine and therefore
left to the reader.
\blemma[Initial Condition for Standard Regular Ensembles]
\label{lem:initstandard}
Consider transmission over the channel BEC$(n, n \epsilon)$ using
a random element of $\eldpc n {x^{\dl-1}} {x^{\dr-1}}$. 
Consider the residual graph (after reception of the transmitted word) and
let $\prob_{\rm init}(s,t)$ denote the distribution of
check nodes of degree one and of degree at least two, respectively.
Then
\[
\prob_{\rm init}(s,t) = \prob_{\rm Gauss}(s,t)
\left(1+O(1/n)\right),
\]
where $\prob_{\rm Gauss}(s,t)$ is a (discrete) Gaussian density 
with mean
\begin{eqnarray*}
\frac{1}{n}\expectation[s] & = & \dl \epsilon (1-\epsilon)^{\dr-1} \, ,\\
\frac{1}{n}\expectation[t] & = & \frac{\dl}{\dr}\left(1 - (1-\epsilon)^{\dr} - \dr \epsilon (1-\epsilon)^{\dr-1} \right) \, ,
\end{eqnarray*}
and covariance
\begin{eqnarray*}
\frac{1}{n}\expectation[s;s] & = & \dl \epsilon \bar{\epsilon}^{\dr-1}(1 - \bar{\epsilon}^{\dr-2}(1 + \epsilon ((\dr-1)\epsilon-1) \dr))\, ,\\
\frac{1}{n}\expectation[s;t] & = & -\dl \epsilon \bar{\epsilon}^{\dr-1} (1 - \bar{\epsilon}^{\dr-2} (1+ \epsilon ((\dr-1)^2 \epsilon-1))), \\
\frac{1}{n}\expectation[t;t] & = & \frac{\dl \bar{\epsilon}^{\dr-1}}{\dr}(1+(\dr-1)\epsilon-\bar{\epsilon}^{\dr-2} (1+\epsilon(2 \dr-3+(\dr-3)(\dr-1)\epsilon+(\dr-1)^3 \epsilon^2)))\, .
\end{eqnarray*}
\elemma

\blemma[Initial Condition for Regular Poisson Ensemble]
A statement analogous to Lemma
\ref{lem:initstandard} holds in the case of Poisson ensembles. For $r=0$ the distribution
of $s$ and $t$ is again a (discrete) Gaussian with mean
\begin{eqnarray*}
\frac{1}{n}\expectation[s] & = & \dl \epsilon\; e^{-\dl\epsilon}\, ,\\
\frac{1}{n}\expectation[t] & = & 1-
e^{-\dl\epsilon}-\dl \epsilon\; e^{-\dl\epsilon}\, ,
\end{eqnarray*}
and covariance
\begin{eqnarray*}
\frac{1}{n}\expectation[s;s] & = & \dl \epsilon \; e^{-\dl\epsilon}-
\dl\epsilon(1-\dl\epsilon+\dl^2\epsilon^2) \; e^{-2\dl\epsilon}\, ,\\
\frac{1}{n}\expectation[s;t] & = & -\dl\epsilon\; e^{-\dl\epsilon}
+\dl\epsilon (1+\dl^2\epsilon^2)\; e^{-2\dl\epsilon}\, ,\\
\frac{1}{n}\expectation[t;t] & = & (1+\dl\epsilon)\, e^{-\dl\epsilon}
-(1+2\dl\epsilon+\dl^2\epsilon^2+\dl^3\epsilon^3)\, e^{-2\dl\epsilon}\; \, .
\end{eqnarray*}
\elemma
Note that, as one would expect, the random variables $(s, t)$ are in general correlated.

\begin{figure}[htp]
\begin{center}
\setlength{\unitlength}{1bp}%
\begin{picture}(0,0)
\includegraphics[scale=1]{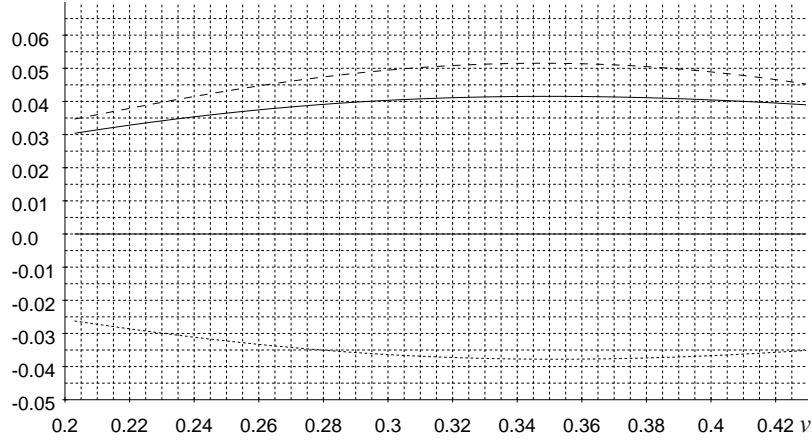}
\end{picture}%
\begin{picture}(320,180)
\put(319, 20){\makebox(0,0){$v$}}
\end{picture}
\end{center}
\caption{\label{fig:varianceevolution} The evolution of $\delta^{(ss)}$ 
(dashed line), $\delta^{(st)}$ (dotted line) and $\delta^{(tt)}$ (solid line) for the
$\eldpc n {x^{2}} {x^{5}}$ ensemble and the choice $\epsilon=\epsilon^* \approx 0.42944$.}
\end{figure}

We can now solve equations~(\ref{GeneralDensityEvolution}) and
(\ref{GeneralCovarianceEvolution}). This allows us to track the evolution of the probability 
distribution of $s$ and $t$ as $v$ decreases from
$n\epsilon$ to $0$, assuming that the $s=0$ plane was not hit earlier. 
\bexample[$(3, 6)$-Ensemble]
Figure \ref{fig:varianceevolution} shows the evolution 
of $\delta^{(ss)}$, $\delta^{(st)}$, $\delta^{(tt)}$ for
the $\eldpc n {x^{2}} {x^{5}}$ ensemble for the choice $\epsilon=\epsilon^* \approx 0.42944$.
Notice that the variances of $s$ and $t$ can actually {\em shrink}
as the decoding process evolves. This is an effect of the 
term in square brackets in equation (\ref{GeneralCovarianceEvolution}).
In particular the variance  shrinks to 0 at $\nu=0$ if $\epsilon$ is low 
enough (whenever decoding is successful with high probability).
\eexample
Finally, the parameter $\alpha$ is given by equation (\ref{equ:alphaformula}), 
where the first factor can be computed as in equation (\ref{equ:criticalpoint}).
%

In Table (\ref{tab:ParametersRegular}) we report the values of $\epsilon^*$, $\alpha$, and $\beta$ 
for a few regular standard ensembles. Further explanations concerning 
the parameter $\beta$ are provided in Section~\ref{sec:shiftparameters}.

\begin{table}
\begin{center}
\begin{tabular}{|c|c||c|c|c|} \hline
$\dl$ & $\dr$ & $\epsilon^*$ & $\alpha$ & $\beta/\Omega$  \\ \hline \hline
3 & 4 & $0.6473$ & $0.260115$ & $0.593632$ \\ \hline
3 & 5 & $0.5176$ & $0.263814$ & $0.616196$ \\ \hline
3 & 6 & $0.4294$ & $0.249869$ & $0.616949$ \\ \hline
4 & 5 & $0.6001$ & $0.241125$ & $0.571617$ \\ \hline
4 & 6 & $0.5061$ & $0.246776$ & $0.574356$ \\ \hline
5 & 6 & $0.5510$ & $0.228362$ & $0.559688$ \\ \hline
6 & 7 & $0.5079$ & $0.280781$ & $0.547797$ \\ \hline
6 & 12 & $0.3075$ & $0.170218$ & $0.506326$ \\ \hline
\end{tabular}
\label{tab:ParametersRegular}
\end{center}
\caption{Thresholds and scaling parameters for some regular standard ensembles.
The shift parameter is given
as $\beta/\Omega$ where $\Omega$ is the universal constant stated in equation (\ref{equ:omega})
whose numerical value is very close to 1.}
\end{table}


%
%

The computation of the scaling parameters 
$\alpha=\alpha(\dl, r)$ and $\beta=\beta(\dl, r)$ for the Poisson case
are made easier by the following pleasing relationship. 
\blemma [Scaling of Erasure Probability for Poisson Ensembles]
Consider transmission over BEC$(n, n \epsilon)$ using elements from
the regular Poisson ensemble $\eldpc n {x^{\dl-1}} r$.
For $\dl$ fixed and $(n, r, \epsilon)$ and $(n', r', \epsilon')$ such that
$n \epsilon = n' \epsilon'$ and $(1-r) n = (1-r') n'$,
\begin{align*}
\expectation_{\eldpc n {x^{\dl-1}} r}[\prob_{\text{B}}(\graph, n \epsilon)]  
& =\expectation_{\eldpc {n'} {x^{\dl-1}} {r'}}[\prob_{\text{B}}(\graph, n' \epsilon')]\, , \\
n \expectation_{\eldpc n {x^{\dl-1}} r}[\prob_{\text{b}}(\graph, n \epsilon)]
& =n' \expectation_{\eldpc {n'} {x^{\dl-1}} {r'}}[\prob_{\text{b}}(\graph, n' \epsilon')]\,. 
\end{align*}
\elemma
\begin{proof}
We start with the statement regarding the block erasure probability.
Compare transmission
over BEC$(n, n \epsilon)$ using elements from $\eldpc {n} {x^{\dl-1}} {r}$
to transmission over BEC$(n', n' \epsilon')$ using elements from
$\eldpc {n'} {x^{\dl-1}} {r'}$. 
The condition $n \epsilon = n' \epsilon'$ implies
that the number of erased bits is the same in both cases.
Decoding fails if these erased bits contain a stopping set.
The condition $(1-r) n = (1-r') n'$ implies that
the two ensembles have the same number of check nodes. Together with the
fact that $\dl$ is the same in both cases (and therefore the involved number
of edges is the same) this shows that the erasure probability is the same.

The proof regarding the bit erasure probability is almost identical.
Both decoders get stuck in identical constellations. The factor $n$
takes into account what fraction of the overall codeword this constellation is.
\end{proof}

If we combine the above relationship with the general form of the scaling
law, cf. equations (\ref{equ:simplscaling}) and (\ref{equ:scaling}) as well as 
Lemma \ref{UnconditionalLemma}, we get the following scaling 
relations.
\blemma[Scaling of Scaling Parameters]
Consider transmission over  BEC$(n, n \epsilon)$ using
elements of the Poisson ensemble $\eldpc n {x^{\dl-1}} r$
with threshold $\epsilon^*(\dl, r)$.
Assume that the scaling (\ref{equ:scaling}) holds and
let $\alpha(\dl, r)$ and $\beta(\dl,r)$ denote the corresponding 
variance and shift parameters. 
Then
\begin{eqnarray}
\epsilon^*(\dl, r') & = & \epsilon^*(\dl, r) \, \frac{1-r'}{1-r}\, ,
\label{equ:epsrelation}\\
\alpha(\dl, r')& = & \alpha(\dl, r) \, \left(\frac{1-r'}{1-r}\right)^{1/2}
\, ,\label{equ:alpharelation}\\
\beta(\dl, r') & = & \beta(\dl, r) \, \left(\frac{1-r'}{1-r}\right)^{1/3}\, .
\label{equ:betarelation}
\end{eqnarray}
\elemma
\begin{proof}The proof is elementary and we leave it to the reader.
We note that in order to prove (\ref{equ:epsrelation}) and (\ref{equ:alpharelation})
only the simplified form of the scaling law (\ref{equ:simplscaling}) is required
as hypothesis and that this scaling law is proved
in Lemma \ref{UnconditionalLemma}.
\end{proof}

From the above observations it follows that we have to determine the parameters
$\epsilon^*(\dl, r)$, $\alpha(\dl, r)$ and $\beta(\dl, r)$ only for one rate $r$. 
This is the reason why so far we have only considered Poisson ensembles of zero rate.
Our results will depend only on 
$\dl$. Relations (\ref{equ:epsrelation})-(\ref{equ:betarelation})
can be used to reintroduce the dependence upon $r$.

\begin{table}
\begin{center}
\begin{tabular}{|c||c|c|c|c|} \hline
$\dl$ & $\epsilon^*$ & $\alpha$ & $\beta/\Omega$ \\ \hline \hline
3 & $0.818469$ & $0.497867$ & $0.964528$  \\ \hline
4 & $0.772280$ & $0.409321$ & $0.827849$  \\ \hline
5 & $0.701780$ & $0.375892$ & $0.760593$  \\ \hline
6 & $0.637081$ & $0.354574$ & $0.713490$  \\ \hline
7 & $0.581775$ & $0.337788$ & $0.676647$  \\ \hline
8 & $0.534997$ & $0.323501$ & $0.646335$  \\ \hline
9 & $0.495255$ & $0.310948$ & $0.620646$  \\ \hline
10 &$0.461197$ & $0.299739$ & $0.598429$  \\ \hline
\end{tabular}
\end{center}
\caption{Thresholds and scaling parameters for some Poisson ensembles
$\eldpc {n} {x^{\dl-1}} {r}$. Note that these parameters assume that $r=0$. 
Parameters for a generic rate can be obtained from these parameters through 
equations.~(\ref{equ:epsrelation})-(\ref{equ:betarelation}). The shift parameter is given
as $\beta/\Omega$ where $\Omega$ is the universal constant stated in (\ref{equ:omega})
whose numerical value is very close to 1.}
\label{tab:PoissonParameters}
\end{table}

\section{Computation of the Shift Parameter}
\label{sec:shiftparameters}

In this section we explain in greater detail the arguments for
Conjecture \ref{RefinedConjecture}, and the procedure for computing the shift parameter 
$\beta$. As in the previous section, we shall first discuss
this issue in an abstract setting, cf. Section \ref{sec:ShiftGeneral}. 
The general procedure will then be applied to regular standard and Poisson ensembles 
in Section \ref{sec:ShiftExamples}.

%
%
\subsection{The General Approach}
\label{sec:ShiftGeneral}

Let us reconsider the setting of Section \ref{sec:GeneralCovarianceEvolution},
i.e., a family of Markov chains $X_{n,0},X_{n,1},\dots,X_{n,t},
\dots$  taking values in ${\mathbb Z}^{d+1}$ and
parametrized  by the (large) integer $n$. As before we will drop in the sequel
the subscript $n$ to mitigate the notational burden.
Throughout this section we shall assume the hypotheses of Proposition
\ref{VarianceEvolutionProp} to be fulfilled.
Unlike in Section~\ref{sec:GeneralCovarianceEvolution},
we are interested in paths $X_0^t \equiv \{X_{0},X_{1},\dots,X_{t}\}$ 
which are confined to the `half space':
\begin{eqnarray}
\Hp \equiv \{ x= (x^{(0)}, \dots, x^{(d)})\in {\mathbb Z}^{d+1}
:\;\; x^{(0)}> 0\}\, .
\end{eqnarray}
We would like to estimate the `survival' probability
\begin{eqnarray}
P_t \equiv\prob (X_0^t\subseteq {\mathbb H}_+)\, .
\end{eqnarray}
Notice that $P_t$ depends implicitly on the initial condition 
$X_0 = x_0\in \Hp$. 
The coordinate $X^{(0)}_t$ should be thought 
as (an abstraction of) the number $s$ of degree-one check nodes 
in the analysis of iterative decoding, cf. Section \ref{sec:general}.
The survival probability $P_t$ is therefore the probability
of not having encountered a stopping set after $t$ steps of the decoding 
process.
We are interested in a time window of length $O(n)$. Without
loss of generality we may fix $\tau_{\rm max}>0$ and
consider $t\in \{0,\dots,t_{\rm max}\}$ with $t_{\rm max} = \lfloor
n\tau_{\rm max}\rfloor$.

We shall denote by $\oz(\tau)$ the `critical trajectory', i.e. a
solution of   the density evolution equations 
(\ref{GeneralDensityEvolution}),
such that $\oz^{(0)}(\tau^*) = 0$, and  $\oz^{(0)}(\tau) > 0$
for any $\tau\in [0,\tau_{\rm max}]$, $\tau\neq\tau^*$.
We call $z_0 = \oz(0)$ the corresponding initial condition.
In order to make contact with the application to
iterative decoding, we shall make the following 
assumptions.
\begin{enumerate}
\item[A.] As $n\to\infty$, we have $x_0 = n\, z_0 + \sqrt{n}\, z_1+ O(1)$,
with $z_1\in{\mathbb R}^{d+1}$ independent of $n$. This
corresponds to the erasure probability $\epsilon$ being in 
the critical window $\epsilon^*-\epsilon = O(n^{-1/2})$.
\item[B.] Let $\oz_{u}(\tau)$, $u\in{\mathbb R}^{d+1}$, be a `perturbed'
critical trajectory obtained by solving the density evolution equations 
(\ref{GeneralDensityEvolution}) with initial condition 
$\oz_{u}(\tau^*) =\oz(\tau^*)+u$. As for the critical trajectory,
we consider this solution in the interval $[0,\tau_{\rm max}]$
and take $u$ such that $|u|<\ve$ with $\ve$ small enough. 
We assume that there exist a positive $u$-independent constant $\kappa_1$,
and a function $u\mapsto a(u)$ such that
\[
\oz^{(0)}_u(\tau)-\oz^{(0)}_u(\tau^*)
\ge a(u)(\tau-\tau^*)+\kappa_1 (\tau-\tau^*)^2
\]
for any $\tau \in [0,\tau_{\max}]$. 
\item[C.] We finally  assume that $a(u)$
can be chosen in such a way that $|a(u)|< \kappa_2 |u|$ for some 
positive constant $\kappa_2$.
\end{enumerate}  
Notice that the assumptions B and C above can be easily checked on the
`continuum' transition rates $\Wh(\Delta|z)$ introduced in 
Sec. \ref{sec:GeneralCovarianceEvolution}. The situation considered
here mimics the one found in iterative decoding of unconditionally stable 
ensembles.

Consider the survival probability
$P_{t_{\rm max}}$ at the `latest' time. As we have seen in 
Section \ref{sec:GeneralCovarianceEvolution}, most of the trajectories 
$X_0^{t_{\rm max}}$ are concentrated within $\sqrt{n}$ around 
$n\oz(t/n)$. Therefore the absolute minimum of $X_t^{(0)}$
in the interval $\{0,\dots, t_{\rm max}\}$ will
be realized for a $t$ `close' to $n \tau^*$. If this absolute minimum is 
positive,
the corresponding trajectory contributes to $P_{t_{\rm max}}$, otherwise it does not.

In order to formalize this argument, fix $t^*= \lfloor n\tau^*\rfloor$. 
Then
\begin{eqnarray}
P_{t_{\rm max}} = \sum_{x\in \Hp\ } 
\prob(X_{0}^{t_{\rm max}}\subseteq \Hp |X_{t^*} = x)\; 
\prob(X_{t^*} = x) \, .\label{eq:SurvivalRepresentation}
\end{eqnarray}
Thanks to Proposition \ref{VarianceEvolutionProp} we can accurately estimate 
the factor $\prob(X_{t^*} = x)$. 
The term $\prob(X_{0}^{t_{\rm max}}\subseteq \Hp |X_{t^*} = x)$ is the probability
that the global minimum of $X^{(0)}_t$, $t\in\{0\dots t_{\rm max}\}$,
is positive conditioned on $X_{t^*}=x$.
\begin{figure}[htp]
\begin{center}
\setlength{\unitlength}{1bp}%
\begin{picture}(0,0)
\epsfig{file=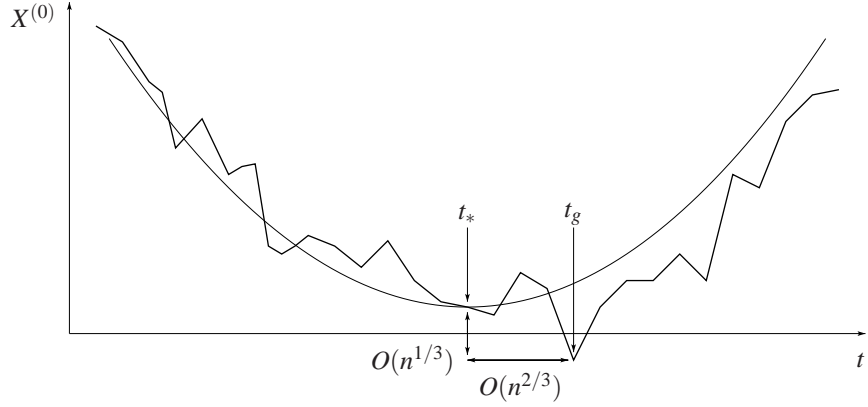,scale=1.0}
\end{picture}%
\begin{picture}(310, 160)
\put(160,75){\makebox(0,0){$t_*$}}
\put(200,75){\makebox(0,0){$t_g$}}
\put(180,10){\makebox(0,0){$O(n^{2/3})$}}
\put(155,20){\makebox(0,0)[r]{$O(n^{1/3})$}}
\put(310,20){\makebox(0,0)[r]{$t$}}
\put(5,150){\makebox(0,0)[r]{$X^{(0)}$}}
\end{picture}
\end{center}
\caption{A pictorial view of decoding trajectories near the critical
point. The type of trajectory depicted here is responsible for 
the shift appearing in the refined scaling form (\ref{equ:scaling}).}
\label{fig:pictorial}
\end{figure}
Let us denote by $t_{\rm g}$ {\em a} `time'  for which the global minimum 
is realized. More precisely, $t_{\rm g}\in\{0\dots t_{\rm max}\}$
is {\em a} random variable such that $X^{(0)}_{t_{\rm g}}\le X^{(0)}_{t}$
for all $t\in \{0\dots t_{\rm max}\}$. 
Call $\oz_X(\tau)$ the perturbed critical trajectory defined above with
perturbation vector $u = X_{t^*}/n-\oz(\tau^*)$. 
In other words, we perturbe the critical trajectory by an 
$O(1/\sqrt{n})$ amount in order to match it to the particular (finite $n$)
realization of the Markov process we are dealing with within the critical 
region.
Concentration arguments, analogous to the ones used to 
prove the point I of Proposition \ref{VarianceEvolutionProp}, imply that,
for a given $t$:
\begin{eqnarray*}
\prob\left\{|X_t-n\oz_X(t/n)|\ge \delta \sqrt{|t-t^*|}\right\}\le \Omega_1 \, 
e^{-\Omega_2\, \delta^2}\, ,
\end{eqnarray*}
for some positive constants $\Omega_1$ and $\Omega_2$ (as before we
use this symbols to denote generic constants which are proven to exist
independent of $n$). 
In fact a stronger condition holds true: 
by Doob's maximal inequality \cite[p.~227]{HDRR98}, for $T$ fixed
\begin{eqnarray}
\prob\left\{ \max_{|t-t^*| \leq T} \,
\;|X_t-n\oz_X(t/n)|\ge \delta \sqrt{T}\right\}
\le \Omega_1 \, e^{-\Omega_2\,\delta^2}\, ,\label{eq:DoobMaximal}
\end{eqnarray}
for some (possibly different)  constants $\Omega_1$ and $\Omega_2$.
Using this fact we can prove an useful result:
\blemma
\label{MinimumLemma} 
Assume the same hypotheses as in Lemma \ref{VarianceEvolutionProp} 
plus A, B and C above.
Let $t_{\rm g}$ be a time at which the absolute minimum 
of $X^{(0)}_t$ is realized, for $t\in\{0\dots t_{\rm max}\}$. 
Then there exist positive 
constants $\Omega_1$, $\Omega_2$ and $\delta_0$, and a function $n_0(\delta)$ 
such that, for any $\delta>\delta_0$ and $n>n_0(\delta)$
\begin{eqnarray}
\hspace{-0.3cm}\prob\left\{ |t_{\rm g}-t^*| \le \delta^{2/3}\, n^{2/3} ,\;
X^{(0)}_{t_g} \ge X^{(0)}_{t^*}-\delta^{4/3}\, n^{1/3}\right\}\ge
1-\Omega_1\,\exp[-\Omega_2\, \delta^2]\, .
\label{eq:MinimumLocation}
\end{eqnarray}
\elemma
The proof is deferred to Appendix \ref{app:Lemma5.1}. The content of this
lemma is illustrated in Fig.~\ref{fig:pictorial}.

The above result implies that corrections to the simplified scaling 
of Lemma \ref{UnconditionalLemma} can be estimated through a two step 
procedure. In a nutshell: $(i)$ Compute the probability for $X_{t^*}^{(0)}$
to be of order $n^{1/3}$; $(ii)$ Evaluate the probability for 
$X^{(0)}_{t_{\rm g}}$ to be positive, conditioned on a given  
$X_{t^*}^{(0)}$ of order $n^{1/3}$.

\subsubsection{Distribution of $X_{t^*}$}
The simplified scaling
form, cf. Lemma \ref{UnconditionalLemma}, was obtained by approximating the 
first factor in equation (\ref{eq:SurvivalRepresentation})
by 1. The leading correction to this approximation comes from trajectories 
such that  $X_{t^*}^{(0)} =O(n^{1/3})$. 
Because of Proposition \ref{VarianceEvolutionProp},
the probability distribution of $X_{t^*}^{(0)}$ (second factor) is well
approximated by a Gaussian with center at $O(\sqrt{n})$ and variance of
order $n$. The probability of having  $X_{t^*}^{(0)}= O(n^{1/3})$
is therefore of order $n^{1/3}\cdot n^{-1/2}=n^{-1/6}$. This
explains why the correction term in the refined scaling form 
(\ref{equ:scaling}) is of order $n^{-1/6}$.

This argument can be made more precise by rewriting 
equation~(\ref{eq:SurvivalRepresentation}) as
\begin{eqnarray}
P_{t_{\rm max}} = \prob(X^{(0)}_{t^*}>0) - \sum_{x\in\Hp}
\prob(X_{t_{\rm g}}^{(0)}<0 |X_{t^*} = x)\; 
\prob(X_{t^*} = x)\, .
\end{eqnarray}
The first term corresponds to the simplified scaling form. We shall 
hereafter focus on the second one, $P_{\rm corr} \equiv \prob(X^{(0)}_{t^*}>0)
-P_{t_{\rm max}}$. Notice that $\prob(X_{t_{\rm g}}^{(0)}<0 |X_{t^*} = x)$
varies much more rapidly (on a scale of order $n^{1/3}$) in $x^{(0)}$ 
than in the other coordinates (on a scale of order $n$).
It is therefore useful to introduce the notation
$\vx = (x^{(1)}\dots x^{(d)})$ (and analogously $\vX$ and $\vz$)
which distinguish explicitly the last $d$ coordinates of $x$. Since 
$\prob(X_{t^*} = x)$ varies on a scale $n^{1/2}$, we can safely
approximate it by setting the coordinate $x^{(0)}$ to 0:
\begin{eqnarray*}
P_{\rm corr} = \sum_{\vx}\left\{\sum_{x^{(0)}>0} 
\prob\left(X_{t_{\rm g}}^{(0)}<0 |X_{t^*} = (x^{(0)}, \vx)\right)\right\}
\prob\left(X_{t^*} = (0, \vx)\right)\, (1+O(n^{-1/6}))\, .
\end{eqnarray*}
The term in curly brackets depends on $\vx$ only through the 
transition coefficients in a neighborhood of $\vx$ and varies therefore on a 
scale of order $n$. This point will be discussed in detail in the next
section. On the contrary $\prob \left( X_{t^*} = (0, \vx) \right)$ 
is peaked around $n\vz(t^*/n)$ with a width of order $\sqrt{n}$.
Therefore
\begin{eqnarray}
P_{\rm corr} = \sum_{x^{(0)}>0} 
\prob\left( X_{t_{\rm g}}^{(0)}<0 |X_{t^*} = (x^{(0)}, n\vz(\tau^*))\right)
\prob \left( X^{(0)}_{t^*} = 0 \right)\, (1+O(n^{-1/6}))\, ,\nonumber\\
\label{eq:ProductFormula}
\end{eqnarray}
where we recall that $\vz(\tau^*)$  denotes the last $d$ coordinates of
the critical point.
The second factor can be evaluated easily using density and covariance evolution. 
Let us consider the application to iterative decoding (here $X^{(0)} \equiv s$).
Note that at the critical point and within the critical window 
$X^{(0)}$ is Gaussian with mean $\frac{\partial \sigma}{\partial \epsilon} (\epsilon-\epsilon^*) n$
and variance $\delta_{\sigma\sigma} n$.
We therefore have
\begin{eqnarray*}
\prob \left( X^{(0)}_{t^*} = 0 \right) = \frac{1}{\frac{\partial \sigma}{\partial \epsilon} \sqrt{2\pi n\alpha^2}}\,
\exp\left\{-\frac{n(\epsilon^*-\epsilon)^2}{2\alpha^2}\right\}
(1+O(n^{-1/2}))\, .
\end{eqnarray*}
This formula can indeed be guessed without any computation at all. 
The probability of  $X^{(0)}_{t^*} = 0$
must be in fact proportional to the derivative of the probability
of having  $X^{(0)}_{t^*} \le 0$, which is given by 
equation~(\ref{equ:simplscaling}) within the  critical window.

\subsubsection{Distribution of the Global Minimum}
We are left with the task of estimating the first factor in 
equation~(\ref{eq:ProductFormula}), and more generally the probability distribution
of $X^{(0)}_{t_{\rm g}}$ conditioned on $X_{t^*}$. 
Lemma \ref{MinimumLemma} is, once again, quite helpful.
The difference $|t_{\rm g}-t^*|$ is small on the scale $n$ on which the
transition rates are state-dependent. This suggests that the leading correction
to the simplified scaling depends on the transition rates only
through their behavior at the critical point $\oz(\tau^*)$. On the
other hand,  $|t_{\rm g}-t^*|$ is large on the scale $O(1)$ of a 
single step. We can therefore hope to compute the leading correction within a 
`continuum' approach.

More precisely, define the rescaled trajectory $u(\cdot)\in{\mathbb R}^{d+1}$ 
by taking
\begin{eqnarray}
u^{(0)}(n^{-2/3}(t-t^*)) &\equiv & n^{-1/3} X^{(0)}_t\, ,
\label{equ:Rescaling1}\\
u^{(i)}(n^{-2/3}(t-t^*)) &\equiv & n^{-2/3} (X^{(i)}_t-X^{(i)}_{t^*})\, i=1, \cdots, d, 
\label{equ:Rescaling2}
\end{eqnarray}
for integers $t$  such that $|t-t^*|\le \theta_{\rm MAX} n^{2/3}$,
and interpolating linearly among these points. A textbook result
in the theory of stochastic processes \cite{VaradhanNotes} implies the 
following lemma.
\blemma\label{DiffusionLemma} 
Let $X$ be distributed as above under the condition 
$X_{t^*} = (n^{1/3}\zeta,n\vz(\tau^*))$. The process $u(\cdot)$ defined 
in equations (\ref{equ:Rescaling1}) and (\ref{equ:Rescaling2}) converges as
$n\to\infty$ to a diffusion process with generator:
\begin{eqnarray}
{\cal L}_d = -\left(\sum_{i=1}^d \omega^*_i u^{(i)}\right)
\frac{\partial\phantom{u^{(0)}}}
{\partial u^{(0)}} - \sum_{i=1}^{d}f^{(i)}_*\frac{\partial\phantom{u^{(i)}}}
{\partial u^{(i)}} + \frac{1}{2} f^{(00)}_* \frac{\partial^2\phantom{u^{(0)}}}
{\partial (u^{(0)})^2}\, ,\label{eq:Generator}
\end{eqnarray}
conditioned on $u^{(0)}(0) = \zeta$, and $\vu(0) = \vec{0}$. In the
above formula we used the notation
\begin{eqnarray*}
f^{(i)}_* = \fh^{(i)}(\oz(\tau^*)), \;\;\;\;\; f^{(ij)}_* = \fh^{(ij)}(\oz(\tau^*))
, \;\;\;\; \omega^*_i = 
\left.\frac{\partial \fh^{(0)}}{\partial z_i}\right|_{\oz(\tau^*)}.
\end{eqnarray*}
\elemma
In order not to burden the presentation, the proof of this statement 
is postponed to App. \ref{sec:diffusionproof}.
Notice that the only role of $\theta_{\rm MAX}$ in the above lemma
is to assure that $u(\theta)$ stays within a finite neighborhood
of $u(0)$ with high probability. We want to use the process $u(\theta)$  in
order to compute the second factor in equation~(\ref{eq:ProductFormula})
and therefore the distribution of the absolute minimum of $u(\theta)$.
Let us call $\theta_{\rm g}$ the location of the minimum.
Lemma \ref{MinimumLemma} implies that $|\theta_{\rm g}|<\delta^{4/3}$
with probability at least $1-\Omega_1\exp(-\Omega_2\delta^2)$.
We can therefore safely let $\theta_{\rm MAX}\to\infty$ and consider the 
diffusion process defined above for $\theta\in (-\infty,+\infty)$.

Notice that only the first derivative with respect to the coordinates
$u^{(1)},\dots,u^{(d)}$ appears in equation (\ref{eq:Generator}). The process
$\vu(\theta)$ is therefore deterministic: $u^{(i)}(\theta) = f^{(i)}_*\theta$
for $i=1,\dots,d$. We can substitute this behavior in equation (\ref{eq:Generator})
and deduce that $u^{(0)}(\theta)$ is a time-dependent diffusion process 
with generator
\begin{eqnarray}
{\cal L}_0(\theta) = - \left(\sum_{i=1}^d\omega_i^* f^{(i)}_*\right)
\theta\, \frac{\partial\phantom{u^{(0)}}}{\partial u^{(0)}} +\frac{1}{2} f^{(00)}_*
\frac{\partial \phantom{u_0^2}}{\partial (u^{(0)})^2}.
\end{eqnarray}
It is convenient to rescale $u^{(0)}$ and $\theta$ in order to reduce the 
above generator to a standard form:
\begin{eqnarray}
\oth = (f^{(00)}_*)^{-1/3}\left(\sum_{i=1}^d\omega_i^* f^{(i)}_*\right)^{2/3}
\theta\, ,\;\;\;\;\;\;
w = (f^{(00)}_*)^{-2/3}\left(\sum_{i=1}^d\omega_i^* f^{(i)}_*\right)^{1/3} u^{(0)}\, .
\end{eqnarray}
The generator for $w(\oth)$ has now the form (we keep the same name with 
an abuse of notation)
\begin{eqnarray}
{\cal L}_0(\bar{\theta}) = - \bar{\theta}\, \frac{\partial\phantom{w}}{\partial w} +
\frac{1}{2}\frac{\partial \phantom{w^2}}{\partial w^2}\, .
\end{eqnarray}
A little thought shows that this is equivalent to saying that 
$w(\oth) = w(0) +\oth^2/2 + B(\oth)$ with $B(\oth)$ a two-sided 
standard Brownian motion with $B(0)=0$. The problem of computing 
the distribution of the global minimum of such a process has
been solved in \cite{Gro89}. Adapting the results of this paper we find
\begin{eqnarray}
\prob \left( w(\oth_{\rm g}) - w(0)<-z \right)  = 1-K(z)^2\, ,
\end{eqnarray}
where
\begin{eqnarray}
K(z) = \frac{1}{2}\int \frac{
\Ai (iy)\Bi(2^{1/3}z+iy) - \Ai (2^{1/3}z+iy)\Bi(iy)
}{\Ai(i y)} \; dy\, .
\end{eqnarray}
with $\Ai(\cdot)$ and $\Bi(\cdot)$ the Airy functions defined in \cite{ASt64}.

Putting everything together we get our final result
\begin{eqnarray*}
\sum_{x^{(0)}>0} 
\prob\left(X_{t_{\rm g}}^{(0)}<0 |X_{t^*} = (x^{(0)}, n\vz(t^*/n))\right) = 
n^{1/3} \Omega \; (f^{(00)}_*)^{2/3}
\left(\sum_{i=1}^d\omega_i^* f^{(i)}_*\right)^{-1/3}\!\!\!\!\!\!
\!\!\!\!\!\!\!\! (1+o(1))\, ,\label{eq:FinalGeneralShift}
\end{eqnarray*}
with
\begin{eqnarray}
\Omega \equiv \int_0^{\infty}\!\! [1-K(z)^2]\; dz\, . \label{equ:omega}
\end{eqnarray}
A numerical computation yields $\Omega = 1.00(1)$. 
%
%
\subsection{Application to Regular Standard and Poisson Ensembles}
\label{sec:ShiftExamples}

There is one important difficulty in applying the general scheme 
explained above to iterative decoding: the 
Markov process is not defined for $s<0$. Recall that $s$ corresponds,
in this context, to the `critical' variable $X^{(0)}_t$.
On the other hand, both the drift and diffusion coefficients $\fh^{(i)}(\cdot)$
and $\fh^{(ij)}(\cdot)$ can be continued analytically through the $s=0$ 
plane. Since the final result (\ref{eq:FinalGeneralShift}) depends on the 
transition rates only through these quantities, we are quite confident that 
it remains correct also for iterative decoding applications.

\bconjecture[Shift Parameter for Regular Standard Ensembles]
Consider the regular standard ensemble $\eldpc n {x^{\dl-1}} {x^{\dr-1}}$ or
the regular Poisson ensemble $\eldpc n {x^{\dl-1}} r$. Then
\begin{equation}
\label{equ:shiftparameter}
\beta/\Omega = - (f^{(\sigma\sigma)})^{2/3}
\left[-\frac{\partial f^{(\sigma)}}{\partial\nu}+
\frac{\partial f^{(\sigma)}}{\partial\tau} f^{(\tau)}
\right]^{-1/3}
\left( \frac{\partial\sigma}{\partial\epsilon} \right)^{-1}.
\end{equation}
For the regular standard ensemble $\eldpc n {x^{\dl-1}} {x^{\dr-1}}$ define
\begin{align*}
g(x) & = \frac{\sum_{i=2}^{\dr} \binom{\dr}{i} x^i \bar{x}^{\dr-i} i}{\sum_{i=2}^{\dr} \binom{\dr}{i} x^i \bar{x}^{\dr-i}}, & 
h(x) & =  (\dl-1) \frac{2 \binom{\dr}{2} x^i \bar{x}^{\dr-i}}{\sum_{i=2}^{\dr} \binom{\dr}{i} x^i \bar{x}^{\dr-i} i}
\end{align*}
Then
\begin{eqnarray*}
\beta/\Omega & = & \left( \frac{\partial\sigma}{\partial\epsilon} \right)^{-1}
\left( \frac{\dl-2}{\dl-1} \right)^{2/3}\left(\frac{h'(z) g(z)-\dl h'(z)}{\tau g'(z)}\right)^{-1/3},
\end{eqnarray*}
where $z=\epsilon x^{\dl-1}$ and all parameters are taken at the critical point.
\econjecture

The generic equation (\ref{equ:shiftparameter}) follows directly from 
equation (\ref{eq:FinalGeneralShift}), applied to the iterative decoding setting. 
For regular standard ensembles these expressions can be made somewhat more 
explicit. First we note that at the critical point $f^{(\sigma)} = -1 - f^{(\tau)}$
since with probability approaching one (as $n$ tends to infinity) the variable node
which is pealed off has (only) one check node of degree one attached to it.\footnote{
This is true since at this point the number of degree-one check nodes is sublinear.}
Since $f^{(\sigma)}=0$ at the critical point it follows that $f^{(\tau)} = -1$.
Using again the relationship $f^{(\sigma)} = -1 - f^{(\tau)}$ some calculations show
that $f^{(\sigma\sigma)} = \frac{\dl-2}{\dl-1}$ and that $\frac{\partial f^{(\sigma)}}{\partial \nu}$
and $\frac{\partial f^{(\sigma)}}{\partial \tau}$ can be expressed as indicated.


\bibliographystyle{siam}
\bibliography{lth,lthpub}

\appendix
\section{Covariance Evolution for a General Markov Process}
\label{se:varapp}

In this Section we reconsider the abstract setting of
Section \ref{GeneralCovarianceEvolution} and outline a proof of Proposition 
\ref{VarianceEvolutionProp} under the assumptions 1-3.

\begin{proof}
We start with statement I, whose proof is fairly standard.
Define a Doob's Martingale $\Xh_0,\dots, \Xh_t$,
\begin{eqnarray*}
\Xh_s = \expectation[X^{(i)}_t|X_0,\dots X_s]\, .
\end{eqnarray*}
Note that $\Xh_t = X^{(i)}_t$ and $\Xh_0 = \expectation[X^{(i)}_t]=\oX_t^{(i)}$ so that
\[
\prob\{|X_t^{(i)}-\oX_t^{(i)}|\ge \rho\} = \prob\{|\Xh_t-\Xh_0|\ge \rho\}.
\] 
Therefore, by the Hoeffding-Azuma inequality we will
have proven (\ref{Concentration}) if we can show that $\Xh_0,\dots, \Xh_t$
has bounded differences, more specifically, if we can show that
\[
|\Xh_s - \Xh_{s-1}| \leq \sqrt{\Omega_0}, \;\; 1 \leq s \leq t.
\]
To accomplish this task note that
\begin{eqnarray}
\label{equ:diffbound}
\hspace{-1cm}
|\Xh_s - \Xh_{s-1}|\le \sup_{y,z} |\expectation[X^{(i)}_t|X_0\dots X_{s-1},
X_s = y]-\expectation[X^{(i)}_t|X_0\dots X_{s-1},X_s = z]|\, ,
\end{eqnarray}
where the $\sup$ is taken over all the $y$ and $z$ such that
the trajectories $\{X_0,\dots X_{s-1},X_s = y\}$ 
and  $\{X_0,\dots X_{s-1},X_s = z\}$ have non-vanishing probability.
Consider therefore two realizations of the Markov chain
which coincide up to time $s-1$ but are independent afterwards.
Denote them by $X_0,X_1,\dots$
and $Y_0,Y_1,\dots$, respectively, where by our assumption $X_{\tau}=Y_{\tau}$
for $0 \leq \tau \leq s-1$, but the processes evolve independenly
for $\tau \geq s$.
Since by assumption $|X_s^{(i)}-X_{s-1}^{(i)}| \leq \kappa_1$ and
$|Y_s^{(i)}-Y_{s-1}^{(i)}| \leq \kappa_1$ almost surely it
follows that $|X_s^{(i)}-Y_{s}^{(i)}| \leq 2 \kappa_1$ almost surely.
Define $\delta X_{\tau} = X_{\tau}-Y_{\tau}$ and  
$\delta \oX_{\tau} = \overline{X}_{\tau}-\overline{Y}_{\tau}$. Then we have
for $s \leq \tau < t$
\begin{eqnarray*}
\delta \oX^{(i)}_{\tau+1} & \le & \delta \oX^{(i)}_\tau +
\expectation [|f^{(i)}(X_\tau)-f^{(i)}(Y_\tau)|] \le\nonumber \\
& \le & \delta \oX^{(i)}_{\tau} +\frac{A}{n}\delta \oX^{(i)}_{\tau} +\frac{B}{n}\, .
\end{eqnarray*}
Here we approximated $f^{(i)}(X_\tau)-f^{(i)}(Y_\tau)$ by  $\fh^{(i)}(X_\tau/n)-\fh^{(i)}(Y_\tau/n)$
and then used the fact that $\fh^{(i)}(z)$ has bounded derivative.
By Gronwall's Lemma we now get $|\oX_t^{(i)}-\overline{Y}_t^{(i)}|<\sqrt{\Omega_0}$ 
for some suitable constant $\Omega_0$. Since 
$\oX_t^{(i)}=\expectation[X^{(i)}_t|X_0\dots X_{s-1},X_s = y]$ 
for some particular choice of $y$ (and some fixed ``past'' $X_0\dots X_{s-1}$)
and the equivalent statement is true for $\overline{Y}_t^{(i)}$ it follows
from (\ref{equ:diffbound}) that $|\Xh_s - \Xh_{s-1}| \leq \sqrt{\Omega_0}$. 

Notice that equation (\ref{Concentration}) implies
\begin{eqnarray}
\expectation |X_t-\oX_t|^p \le \alpha_p ( \Omega_0 t)^{p/2}\, ,\label{AzumaEstimate}
\end{eqnarray}
for some\footnote{One has in fact $\alpha_p = p\,\sqrt{\pi/2}\;
\expectation|Z|^{p-1}$ with $Z$ a standard Gaussian variable.} 
positive constants $\alpha_p$. Before passing to the following parts of the 
Proposition, let us notice that not all the assumptions on the
transition rates $\Wh(\Delta|z)$ were used here. It is in fact 
sufficient to assume that the drifts $\fh^{(i)}(z)$ are Lipschitz continuous.

Let us now consider the point II. A simple computation shows that
\begin{eqnarray}
\expectation\, X^{(i)}_{t+1} & = & \expectation\, X^{(i)}_t 
+\expectation\, f^{(i)}(X_t)\, ,\\
\expectation [X^{(i)}_{t+1};X^{(j)}_{t+1}] & = &
\expectation [X^{(i)}_{t};X^{(j)}_{t}] +
\expectation\, f^{(ij)}(X_t) +    
\label{SecondMomentRecursion} \\
&&+ \expectation [ X^{(i)}_t ; f^{(j)}(X_t)] +
\expectation [ f^{(i)}(X_t) ;  X^{(j)}_t ] +
\expectation [ f^{(i)}(X_t) ;  f^{(j)}(X_t) ]\, .\nonumber
\end{eqnarray}
Consider the first of these equations and notice that, approximating 
$f^{(i)}(X_t)$ by $\fh^{(i)}(X_t/n)$ one obtains
\begin{eqnarray}
|\oX^{(i)}_{t+1}-\oX^{(i)}_t-\fh^{(i)}(\oX_t/n)| \le
\frac{A}{n}+|\expectation[\fh^{(i)}(X_t/n)-\fh^{(i)}(\oX_t/n)]|\, .
\label{Bound1}
\end{eqnarray}
Since the second derivative of $\fh^{(i)}(z)$ is bounded, we have the estimate
\begin{eqnarray*}
|\expectation[\fh^{(i)}(X_t/n)-\fh^{(i)}(\oX_t/n)]|& \le & 
\left|\frac{1}{n}\sum_j\left.\frac{\partial\fh^{(i)}}{\partial z_j}
\right|_{\oX_t/n}\!\!\!\expectation[X^{(j)}_t-\oX^{(j)}_t]\right|+
\frac{B}{n^2}\expectation |X_t-\oX_t|^2\le \nonumber\\
&\le & \frac{C}{n}\, .
\end{eqnarray*}
Summing equation (\ref{Bound1}) over $t$, and applying Gronwall's Lemma
we get
\begin{eqnarray}
\left|\frac{1}{n}\oX^{(i)}_t-\oz^{(i)}(t/n)\right| & \le& \frac{A'}{n}
\, .\label{FinalFirstMoment}
\end{eqnarray}
Notice that if we limit ourself to assume Lipschitz continuous 
drift coefficients $\fh^{(i)}(z)$, the same derivation yields a slightly 
weaker result: $|\oX^{(i)}_t/n-\oz^{(i)}(t/n)|\le A'/\sqrt{n}$.

Equation (\ref{SecondMomentEstimate}) is proved from 
(\ref{SecondMomentRecursion}) much in the same way, the crucial input being 
an estimate on $\expectation |X_t-\oX_t|^3$, once again obtained from 
equation (\ref{Concentration}). Here we limit ourselves to sketch how the various 
terms emerges. We start by rewriting equation (\ref{SecondMomentRecursion})
in the form
\begin{eqnarray*}
\hspace{-0.5cm}\Delta^{(ij)}_{t+1} & = & \Delta^{(ij)}_t + \fh^{(ij)}(\oX_t/n)+
\frac{1}{n}\sum_{l=1}^d\left[\Delta_t^{(il)}\left.
\frac{\partial\fh^{(j)}}{\partial z_l}\right|_{\oX_t/n} \!\!\!+ 
\left.\frac{\partial\fh^{(i)}}{\partial z_l}\right|_{\oX_t/n}\!\!\!
\Delta^{(lj)}_t\right]+\, \\
&& +R^{(0)}_{ij} +R^{(1)}_{ij}+R^{(1)}_{ji}+R^{(2)}_{ij} + R^{(2)}_{ji}
+R^{(3)}_{ij}\, ,\nonumber
\end{eqnarray*}
With the remainders listed below
\begin{eqnarray*}
R^{(0)}_{ij} & = & \expectation[ f^{(ij)}(X_t)-\fh^{(ij)}(X_t/n)] + 
\expectation[\fh^{(ij)}(X_t/n) - \fh^{(ij)}(\oX_t/n)]\, ,\\
R^{(1)}_{ij} & = & \expectation[X^{(i)}_t; f^{(j)}(X_t)-\fh^{(j)}(X_t/n)]\, ,\\
R^{(2)}_{ij} & = &\expectation[
X^{(i)}_t; \fh^{(j)}(X_t/n)-\fh^{(j)}(\oX_t/n)-\frac{1}{n}
\sum_{l=1}^d\left.\frac{\partial\fh^{(j)}}{\partial z_l}\right|_{\oX_t/n}
(X^{(l)}_t-\oX^{(l)}_t)]\, ,\\
R^{(3)}_{ij} & = & \expectation[f^{(i)}(X_t);f^{(j)}(X_t)]\, .
\end{eqnarray*}
Each of this terms can be bounded separately as in the derivation of 
Eq. (\ref{FinalFirstMoment}).
Consider for instance $R^{(1)}_{ij}$:
\begin{eqnarray*}
|R^{(1)}_{ij}| & \le & \expectation[X^{(i)}_t;X^{(i)}_t]^{1/2}
 \expectation[f^{(j)}(X_t)-\fh^{(j)}(X_t/n);f^{(j)}(X_t)-\fh^{(j)}(X_t/n)]^{1/2}\le\nonumber\\
&\le & A n^{1/2} \frac{B}{n}\le \frac{C}{\sqrt{n}}\, ,
\end{eqnarray*}
where we used the estimate (\ref{AzumaEstimate}).

Let us finally consider part III of the proposition, as stated in
equation (\ref{GeneratingFunctionEstimate}). It is easy to derive the
following recursion for the generating function:
\begin{eqnarray}
\Lambda_{t+1}(\lambda) &= &\Lambda_t(\lambda) +
\log \Wt(\lambda/\sqrt{n}|\oX_t)-\frac{1}{\sqrt{n}}\lambda\cdot
(X_t-\oX_t)+\nonumber\\
&&+\log\left\{\frac{\expectation[\Wt(\lambda/\sqrt{n}|X_t)
\, e^{\frac{\lambda}{\sqrt{n}}\cdot X_t}]}{\expectation[
\Wt(\lambda/\sqrt{n}|\oX_t)\, e^{\frac{\lambda}{\sqrt{n}}\cdot X_t} ]}\right\}
\, .\label{GeneratingFunctionRecursion}
\end{eqnarray}
Here we defined the jump generating function
\begin{eqnarray*}
\Wt(\lambda|x) \equiv \sum_{\Delta} e^{\lambda\cdot \Delta}\,\,
W(\Delta|x)\, .
\end{eqnarray*}
The proof of equation (\ref{GeneratingFunctionEstimate}) is completed  by 
estimating the various terms in equation (\ref{GeneratingFunctionRecursion})
as follows
\begin{eqnarray*}
\left|\log \Wt(\lambda/\sqrt{n}|\oX_t) - \frac{\lambda}{\sqrt{n}}\cdot
(\oX_{t+1}-\oX_t) - \frac{1}{2n}\sum_{i,j} \fh^{(ij)}(\oX_t/n)
\, \lambda_i\lambda_j \right| & \le & \frac{\Omega_a(\lambda)}{n^{3/2}}\, ,
\nonumber\\
\\
\left|\frac{\expectation[(\Wt(\lambda/\sqrt{n}|X_t)-
\Wt(\lambda/\sqrt{n}|\oX_t))
\, e^{\frac{\lambda}{\sqrt{n}}\cdot X_t}]}{\expectation[
\Wt(\lambda/\sqrt{n}|\oX_t)\, e^{\frac{\lambda}{\sqrt{n}}\cdot X_t} ]}-
\phantom{ -\frac{1}{n^2}\sum_{l=1}^d\left[
\left.\frac{\partial\fh^{(i)}}{\partial z_l}\right|_{\oX_t/n}\right] }
\right. & &\\ 
\left. -\frac{1}{n^2}\sum_{l=1}^d\left[
\left.\frac{\partial\fh^{(i)}}{\partial z_l}\right|_{\oX_t/n}
\!\!\!\Delta^{(lj)}_t + \Delta^{(il)}_t
\left.\frac{\partial\fh^{(j)}}{\partial z_l}\right|_{\oX_t/n}\right]
\right| &\le & \frac{\Omega_b(\lambda)}{n^{3/2}}\, . \nonumber
\end{eqnarray*}
We leave to the reader the pleasure of proving these two last (straightforward)
inequalities.
\end{proof}

\section{Unconditionally Stable Ensembles: Proof of the Scaling Law}
\label{app:proof}

In this Appendix we prove Lemma \ref{UnconditionalLemma}. 
The idea is to regard iterative decoding as a Markov process in the space 
of states\footnote{For the sake of definiteness,
we refer here to the case of regular ensembles: the extension
to general unconditionally stable ensembles being trivial. Also,
we use the subscript $\graph$ for the state coordinates in order
to distinguish them from the time parameters $t$ and $\tau$.} 
$x=(v_{\graph},s_{\graph},t_{\graph})\in{\mathbb Z}^3$. 
The transition rates and the
initial condition for such a process are computed in 
Section \ref{StandardRegularParametersSec}.
As in Sec. \ref{sec:GeneralCovarianceEvolution}, we denote 
by $z=x/n = (\nu_{\graph},\sigma_{\graph},\tau_{\graph})$ 
the normalized state and by 
$\oz(\tau)$ the critical trajectory. This is the solution of the density 
evolution equations (\ref{GeneralDensityEvolution}), such that 
$\oz(\tau_{\rm end}) = (0,0,0)$, corresponding to complete decoding,
$\sigma_{\graph}(\tau^*) =0$ for some $\tau^*\in (0,\tau_{\rm end})$, and
$\sigma_{\graph}(\tau)>0$ for any $\tau\in (0,\tau_{\rm end})$, $\tau\neq\tau^*$.

It would be tempting to use
the general covariance evolution approach provided by Proposition
\ref{VarianceEvolutionProp}. However a simple remark 
prevents us from following this route in the most straightforward fashion.
Proposition \ref{VarianceEvolutionProp} was proved under the assumptions
that the transition rates $\Wh(\Delta|z)$ in the $n\to\infty$ limit become
$C^2({\mathbb R}^{d+1})$  functions of $z$. On the other 
hand, the decoding process is well defined only if $s_{\graph}> 0$, 
and we are interested in trajectories passing close to the $s=0$ plane.
In more concrete terms, Proposition  \ref{VarianceEvolutionProp}
cannot be true when $\oz(\tau)$ is at a distance of order $1/\sqrt{n}$
from the $s_{\graph}=0$ plane. 
The least that will happen is that a part of the Gaussian
density is `cut away'.

As a way to overcome this problem, we introduce a new Markov process
on the same states $x=(v_{\graph},s_{\graph},t_{\graph})$ which is well 
defined for $s_{\graph}\le 0$. 
We extend the transition rates computed in the proof of Lemma
\ref{StandardRegularParametersLemma} to  $s_{\graph}\le 0$ by 
setting $\sigma_{\graph}=0$ there. More precisely we have:
\begin{eqnarray}
\Delta v_{\graph} = -1\, ,\;\;\;\;\;\;\;\;
\Delta s_{\graph} = -u_1+u_2\, ,\;\;\;\;\;\;\;\; 
\Delta t_{\graph} = -u_2\, ,\;\;\;\;\;\;\;\; 
\end{eqnarray}
with $u_1$ and $u_2$  distributed according $w(u_1,u_2)$, see 
equation (\ref{eq:ContinuumTransition}),  where we put $q_1=0$ and 
$q_2 = 2\tau_2/\nu\dl$ and $\tau_2$ is determined as in 
(\ref{equ:degreedistribution}). Notice that the only non-zero
entries of the distribution $w(u_1,u_2)$ in the $s_{\graph}\le 0$ space are therefore
\[
w(1,u_2) = \binom{\dl-1}{u_2} q_2^{u_2}(1-q_2)^{\dl-1-u_2}\, .
\]
Such transition rates do not necessarily correspond to any 
graph process in the $s_{\graph}<0$ plane. However, upon conditioning
on $s_{\graph}>0$ the `extended' process coincides with the original one.
Therefore the probability of not leaving the $s_{\graph}>0$ half-space
(the `survival' probability) can be calculated on the extended process.
Finally, let us notice that the precise form of this extension is immaterial
as long as some requirements are met. Call $W(\Delta|x)$ the transition 
rates of the extended Markov process. We require that:
\begin{itemize}
\item The chain makes finite jumps.
\item The rates are well approximated by their continuum counterpart
$\Wh(\Delta|z)$. As in Sec. \ref{sec:GeneralCovarianceEvolution}
this means that $|W(\Delta|x)-\Wh(\Delta|x/n)|\le \kappa/n$.
\item The continuum transition rates are $C^2$ with bounded derivatives in 
the region $\{ \nu_{\graph}>\ve,\, \sigma_{\graph} >\ve,\, \tau_{\graph}>\ve\}$ for any $\ve>0$.
\item There exist a $\delta>0$ such that the continuum 
drift coefficients are Lipschitz continuous uniformly in 
the region ${\rm Crit}(\delta)\equiv \{z \mbox{ s.t. } |z-\oz(\tau^*)|<
\delta\}$. This means that
$|\fh_i(z)-\fh_i(z')|\le \kappa' |z-z'|$ for some positive $\kappa'$ 
and any pair of points $z, z'\in {\rm Crit}(\delta)$. 
\end{itemize}
These requirements are easily checked on the extension defined above.

Recall from Lemma \ref{UnconditionalLemma} that we are only interested
in decoding errors of size at least $\gamma \nu^*_{\graph}$, where 
$\nu^*_{\graph}:=\nu_{\graph}(\tau^*)$ is the
critical point (measured in terms of the fractional size of the graph) 
and $\gamma$ is any number in $(0, 1)$. In particular $\gamma$ is non-negative but
can be chosen arbitrarily small. For ensembles with $\lambda'(0)=0$ a
simple union bound shows that the decoder will be successful with high
probability once the residual graph is sufficiently small but if $\lambda'(0)>0$
then small deficiencies in the graph can contribute non-negligibly to the error probability.
Therefore, by choosing $\gamma \in (0, 1)$, we ``separate out'' the contributions
to the block error probability which stem from large error events.

Call $P_{\rm end}$ the probability of not hitting the $s_{\graph}=0$ until 
$v_{\graph}=\lfloor n \gamma \nu^* \rfloor$. Fix $\tau_{\rm max}$ so
that $\nu(\tau_{\rm max}) = \gamma \nu^*$. 
Define $P_{t}$ to be the survival probability up to time $t$.
It will be useful to denote by  $P_{t}(x',t')$ the probability 
of surviving up to time $t$ conditioned on having survived up to time $t'$ 
and that the state at time $t'$ is $x'$.

In order to apply Proposition \ref{VarianceEvolutionProp} as far as we can,
we decompose the time up to $t_{\rm max}$ into two
intervals: $\{0,\dots ,t_-^*\}$ and $\{t_-^*+1,\dots,t_{\rm max}\}$. 
The survival probability can be written as
\begin{eqnarray}
P_{t_{\rm max}} & =& \sum_{x} P_{t_{\rm max}}(x,t_-^*)\, 
P(x,t_-^*|x_0,0) \, . \label{TwoTimesSurvival}
\end{eqnarray}
Here $P(x',t'|x,t)$ denotes the probability of arriving in state $x'$
at time $t'$  without hitting the $s_{\graph}=0$ plane, conditined on being in  
state $x$ at time $t$. 
The sum over $x$ runs over the 
$s_{\graph}>0$ half-space. 

Next we chose $t_-^* = \lfloor n(\tau^*-\ve)\rfloor$ 
for some (small) positive number $\ve$.
With
this choice the factor $P(x,t_-^*|x_0,0)$ in the above equation
can be estimated using the covariance evolution
approach and Proposition \ref{VarianceEvolutionProp}. The reason is that 
the trajectories contributing to this factor stay at a distance 
of order $n$ from the $s_{\graph}=0$ apart from some exponentially rare cases. 
We leave to the reader the task of adapting the proof 
of Proposition \ref{VarianceEvolutionProp}.III to this situation.

The first factor in equation (\ref{TwoTimesSurvival}) can not be estimated
through covariance evolution. Fortunately a less refined calculation
is sufficient in this case. In fact the Lipschitz  continuity
of the drift coefficients ensures that, at any time $t>t^*_-$, the
state is within $\delta$ of the density evolution prediction
with probability at least $1-\exp[-\delta^2/2\Omega(t-t_-^*)]$. This 
fact was stressed in the proof of Proposition \ref{VarianceEvolutionProp},
cf. Appendix \ref{se:varapp}.
For any state $x$, consider the solution
$\oz(\tau;x)$ of the density
evolution equations (\ref{GeneralDensityEvolution}) with initial condition
$\oz(t_-^*/n;x) = x/n$. Let $\Ph_{t_{\rm max}}(x,t_-^*)=0$
if  $\oz(\tau;x)$ intersects the $\sigma_{\graph}=0$ plane in the interval 
$[t_-^*/n,\tau_{\rm max}]$ and $\Ph_{t_{\rm max}}(x,t_-^*)=1$ otherwise.
The above concentration result implies that $\Ph_{t_{\rm max}}(x,t_-^*)$
is  a good approximation for $P_{t_{\rm max}}(x,t_-^*)$. 

Let us prove the last statement in the cases in which 
$\oz(\tau;x)$  does not intersect the $\sigma_{\graph}=0$ plane
(and therefore  $\Ph_{t_{\rm max}}(x,t_-^*)=1$).
If $x$ is distributed according to $P(x,t_-^*|x_0,0)$, the trajectory
$\oz(\tau;x)$ will stay at a distance of order $1/\sqrt{n}$ from the
critical one. In particular, its minimum distance from the 
$\sigma_{\graph}=0$ plane will be  $\gamma/\sqrt{n}$ with $\gamma$ of
order 1. 
This minimum will be achieved for $\tau$ close to $\tau_*$ with high 
probability.
We therefore restrict ourselves to an interval
of times $t_-^*<t<t_-^*+nT\ve$ for some fixed number $T>1$, and neglect 
the cases in which the $\sigma_{\graph}$ plane is touched outside this 
interval.
The error implied in substituting $\Ph_{t_{\rm max}}(x,t_-^*)$ with 
$P_{t_{\rm max}}(x,t_-^*)$ is upper bounded by the probability 
that the maximum distance between the actual decoding trajectory and
$\oz(\tau;x)$ in the interval $t_-^*<t<t_-^*+nT\ve$ ($\tau^*-\ve<\tau
<\tau_*+(T-1)\ve$) is larger than $\gamma\sqrt{n}$.
Using the above concentration result with $\delta = \gamma\sqrt{n}$ and 
$t-t_-^*<nT\ve$,
we get
\begin{eqnarray}
|\Ph_{t_{\rm max}}(x,t_-^*) -P_{t_{\rm max}}(x,t_-^*)|\le\exp
\left\{-\frac{\gamma^2}{2\Omega T\ve} \right\}\, .
\label{eq:SecondIntervalEstimate}
\end{eqnarray}
As mentioned above,
under the distribution $P(x,t_-^*|x_0,0)$, both $\gamma$ and $T$ are,
with high probability $O(1)$ (both with respect to $n\to\infty$ and 
$\ve\to 0$).
Therefore the right hand side of equation 
(\ref{eq:SecondIntervalEstimate}) can be made arbitrarily small by taking
$\ve\to 0$.

The last step consists in substituting $\Ph_{t_{\rm max}}(x,t_-^*)$
for $P_{t_{\rm max}}(x,t_-^*)$ and the Gaussian density from covariance 
evolution for $P(x,t_-^*|x_0,0)$ in equation (\ref{TwoTimesSurvival}) 
and letting $n\to\infty$
with $n^{1/2}(\epsilon-\epsilon^*)$ fixed. This yields 
Lemma \ref{UnconditionalLemma} up to corrections of which 
vanish when $\ve\to 0$.

\section{Proof of Lemma \ref{MinimumLemma}}
\label{app:Lemma5.1}

In this Appendix we present a proof of Lemma \ref{MinimumLemma},
making use of Doob's maximal inequality (\ref{eq:DoobMaximal}).
We shall prove that each of the two events considered in 
Eq. (\ref{eq:MinimumLocation}) occurs with probability greater than 
$1-\Omega_1\,\exp[-\Omega_2\, \delta^2]$. This implies the thesis
by a simple union bound, plus a rescaling of the constants $\Omega_1$,
$\Omega_2$.

Let us begin by considering the second event, namely
$X^{(0)}_{t_g} \ge X^{(0)}_{t^*}-\delta^{4/3}\, n^{1/3}$. For sake of 
simplicity we redefine $t_{\rm g}$ to be the position of 
the global minimum of $X^{(0)}_t$ in the 
domain $t>t^*$. The minimum with an unrestricted $t$ can be treated by putting
together the cases  $t>t^*$ and $t<t_*$. It is also useful to define
\begin{eqnarray*}
Y_{t-t_*} := \frac{1}{\kappa_1}(X^{(0)}_t-X^{(0)}_{t^*})\, .
\end{eqnarray*}
Equation (\ref{eq:DoobMaximal}) implies
\begin{eqnarray}
\prob\left\{\min_{0\leq t\leq T}\left[Y_t-\frac{1}{n}\, t^2+
\frac{\kappa_2\delta}{\sqrt{n}}\, t\right]\leq -
\delta\sqrt{T}\right\}\le\Omega_1\, e^{-\Omega_2\delta^2}\, ,
\label{eq:DoobMaximal2}
\end{eqnarray}
where we rescaled the constants $\kappa_2$ and $\Omega_2$.

Let $\{ t_l:l\in{\mathbb Z}\}$ be a non-decreasing sequence of real numbers
with $t_l \to \infty$ as $l\to\infty$ and $t_l = 0$ as $l\to -\infty$. 
A union bound yields
\begin{eqnarray*}
&&\prob\left\{\min_{t\geq 0}\; Y_t\leq -\delta^{4/3}n^{1/3}\right\}\leq
\sum_{l=-\infty}^{+\infty}\prob\left\{\min_{t_l\leq t <t_{l+1}}Y_t
\leq -\delta^{4/3}n^{1/3}\right\}\leq\\
&&\leq \sum_{l=-\infty}^{+\infty}
\prob\left\{\min_{t_l\leq t <t_{l+1}}\left[Y_t-\frac{1}{n}\, 
t^2+\frac{\kappa_2\delta}{\sqrt{n}}\, t\right]\le -\delta^{4/3}n^{1/3} -
\frac{1}{n}\, t_l^2 +\frac{\kappa_2\delta}{\sqrt{n}}\, t_{l+1}\right\}\leq\\
&&\leq \Omega_1\sum_{l=-\infty}^{+\infty}\exp\left\{-\Omega_2
\frac{1}{t_{l+1}}\left(\delta^{4/3}n^{1/3} +
\frac{1}{n}\, t_l^2 -\frac{\kappa_2\delta}{\sqrt{n}}\, t_{l+1}\right)\right\}
\, ,
\end{eqnarray*}
where we used Eq. (\ref{eq:DoobMaximal2}) in the last inequality.
At thin point we choose $t_l = 2^l(n\delta)^{2/3}$. Plugging into
the above expression we get
\begin{eqnarray*}
&&\prob\left\{\min_{t\geq 0}\; Y_t\leq -\delta^{4/3}n^{1/3}\right\}\leq
\Omega_1\sum_{l=-\infty}^{+\infty}\exp\left\{-\frac{\Omega_2\delta^2}{2^{l+1}}
\left(1+2^{2l}-\frac{\kappa_2\delta^{1/3}}{n^{1/6}}2^{l+1}
\right)^2\right\}\, .
\end{eqnarray*}
If $n>n_0(\delta) := (2\kappa_2)^6\delta^2$ we get
\begin{eqnarray*}
&&\prob\left\{\min_{t\geq 0}\; Y_t\leq -\delta^{4/3}n^{1/3}\right\}\leq
\Omega_1\sum_{l=-\infty}^{+\infty}\exp\left\{-\frac{\Omega_2\delta^2}{2^{l+1}}
\left(1+2^{2l}-2^l\right)^2\right\}\, .
\end{eqnarray*}
It is an elementary exercise to show that the right hand side is 
smaller than $\Omega_1'\exp\{-\Omega_2'\delta^2\}$ for some
(eventually different) positive parameters $\Omega_1'$ and $\Omega_2'$
and any $\delta>\delta_0$.

The second part of the proof consists in proving an analogous upper bound
for the probability of having $|t_{\rm g}-t^*|>\delta^{2/3}n^{2/3}$.
In fact the proof proceeds as for the first event. One splits the semi-infinite
interval $t>t^*$ in intervals $[t_l,t_{l+1}[$ with $t_l =  2^l(n\delta)^{2/3}$
and (this time) $l\ge 0$, and then apply Doob's
maximal inequality to each interval. We leave to the reader the pleasure 
of filling the details.

\section{Convergence to diffusion process}
\label{sec:diffusionproof}

In this Appendix we prove Lemma \ref{DiffusionLemma} 
as a straightforward application of the following statement which can be 
found in \cite{VaradhanNotes}.

\begin{theorem}
Let $\{ X_t\}$ be a Markov process with values in $\R^d$ and transition 
probability $\pi_h(x,dy)$, with $0<h\le 1$ and initial condition $X_0=x_0$.
Let $P_h$ be the measure induced on the space of continuous trajectories
$\Omega=C([0,\infty),\R^d)$ by the mapping $X(th)=X_t$ for integer $t$
and interpolating linearly in between. Assume that the limit 
\begin{eqnarray}
\lim_{h\to\infty} \frac{1}{h}\int_{\R^d}\! [\phi(y)-\phi(x)]\;\pi_h(x,dy) = 
({\cal L}\phi)(x)\, ,
\end{eqnarray}
exists uniformly in a compact $K\subseteq \R^d$ for functions 
$\phi\in C^{\infty}(K)$. Assume that the limit has the form
\begin{eqnarray}
({\cal L}\phi)(x) = \frac{1}{2}\sum_{ij}a_{ij}(x)\frac{\partial^2 \phi}
{\partial x_i\partial x_j} + \sum_{i=1}^d b_i(x)\frac{\partial \phi}
{\partial x_i}\, ,
\end{eqnarray}
with continuous and uniformly bounded coefficients $a \equiv \{ a_{ij}(x)\}$
($a$ being a positive definite matrix) and  $b\equiv \{ b_{i}(x)\}$.
Assume finally that the solution of the martingale problem for ${\cal A}$ 
is unique yielding a Markov family of measures $P_x$ on $\Omega$.
Then $\{P_h,x\}$ converges to $\{P_x\}$ as $h\to 0$.
\end{theorem}

The proof of Lemma \ref{DiffusionLemma} proceed then sa follows.
Set $h=n^{-2/3}$ and define the a Markov chain in the variables 
$u_0, \vu$, see Eq. (\ref{equ:Rescaling1}), (\ref{equ:Rescaling1})  
using the transition rates $W(\Delta|x)$ and the initial condition 
$u_0(0) =\zeta$, $\vu(0) = 0$.  One has then just to compute the generator
\begin{eqnarray}
({\cal L}\phi)(u_0,\vu) =  \lim_{n\to\infty} n^{2/3}\sum_{\Delta_0,\vD}
[\phi(u_0+n^{-1/3}\Delta_0,\vu+n^{-2/3}\vD)-f(u_0,\vu)]\cdot\nonumber\\
\phantom{({\cal L}\phi)(u_0,\vu) =  \lim_{n\to\infty} n^{2/3}\sum_{\Delta_0,\vD}\;\;\cdot}
\Wh(\Delta_0,\vD|n^{-2/3}v_0,n^{-1}\vX_{t_*}+n^{-1/3}\vu)\, ,
\end{eqnarray}
where made the subsitution $W(\Delta|x)\to \Wh(\Delta|x/n)$ which 
implies a negligible $O(1/n)$ error. The formula (\ref{eq:Generator})
is easily obtained by Taylor expansion the above equation.

\end{document}